\documentclass[a4paper,11pt,titlepage,oneside]{book} 

\pdfoutput=1
\usepackage[T1]{fontenc}
\usepackage[utf8]{inputenc}

\usepackage{ae}               
\usepackage[pdftex]{graphicx}
\usepackage{vmargin}         
\usepackage{fancyhdr}         
\usepackage{subfigure}
\usepackage[hyphens]{url}
\usepackage{xcolor}
\usepackage{footnote}

\usepackage[absolute,overlay]{textpos}
\usepackage{tikz}
\usepackage[english]{babel}
\usepackage{color}
\usepackage{longtable}
\usepackage[bottom]{footmisc}
\usepackage[section,subsection,subsubsection]{extraplaceins}
\usepackage{multirow}
\usepackage{multicol}	
\usepackage{array}
\newcolumntype{P}[1]{>{\raggedright\arraybackslash}p{#1}}
\usepackage{nomencl}
\usepackage{color}
\usepackage{setspace}

\makenomenclature

\makeglossary

\definecolor{darkred}{rgb}{0.5,0,0}
\definecolor{darkgreen}{rgb}{0,0.8,0}
\definecolor{darkblue}{rgb}{0,0,0.5}

\setmarginsrb
{2.5cm}  
{1.5cm}  
{2.5cm}  
{2cm}   
{20pt}    
{0.25in}  
{9pt}     
{0.3in}

\usepackage{chngcntr}
\counterwithin{figure}{section} 
\counterwithin{table}{section}  

\renewcommand\thetable{\thesection.\arabic{table}}

\usepackage{authblk}

\renewcommand{\thesection}{\arabic{section}}

\newcommand\mytitle{Cloud Usage Patterns: A Formalism for Description of Cloud Usage Scenarios}
\title{Cloud Usage Patterns}

\newcommand\TRnumber{Technical Report: SPEC-RG-2013-001\\Version: 1.0.1}
\newcommand\WGname{SPEC RG Cloud Working Group}

\newcommand\TRdate{May 28, 2013}
\newcommand\TRcentralURL{research.spec.org}
\newcommand\TRrightURL{www.spec.org}

\newcommand\Acknowledgements{The authors would like to thank  Michael Faber who is with Deutsche Bank, Germany, and Jon Curtiss and Diane Mularz who are with the MITRE Corporation, USA, for the fruitful discussions and comments to improve this work.  
}

\newcommand\numAuthors{8} 
\makeatletter
\newcommand\defcase[1]{\@namedef{mycase@\the\numexpr#1\relax}}
\newcommand\putAuthors[1]{\@nameuse{mycase@\the\numexpr#1\relax}}
\makeatother

\defcase{0}{
	\begin{textblock}{8}[0,0](\colsinglecentralX,\rowoneY)
		\centering
		\small{\textbf{No Author}}\\
	\end{textblock}	
}

\defcase{1}{
	\begin{textblock}{8}[0,0](\colsinglecentralX,\rowoneY)
		\centering
		\small{\textbf{\authorOneName}}\\
		\footnotesize
		\authorOneAffil
	\end{textblock}	
}
\defcase{2}{
	\begin{textblock}{\authorCellWidth}[0,0](\colDoubleLeftX,\rowoneY)
		\centering
		\small{\textbf{\authorOneName}}\\
		\footnotesize
		\authorOneAffil
	\end{textblock}	
	\begin{textblock}{\authorCellWidth}[0,0](\colDoubleRightX,\rowoneY)
		\centering
		\small{\textbf{\authorTwoName}}\\
		\footnotesize
		\authorTwoAffil
	\end{textblock}	
}
\defcase{3}{
	\begin{textblock}{\authorCellWidth}[0,0](\coloneX,\rowoneY)
		\centering
		\small{\textbf{\authorOneName}}\\
		\footnotesize
		\authorOneAffil
	\end{textblock}	
	\begin{textblock}{\authorCellWidth}[0,0](\coltwoX,\rowoneY)
		\centering
		\small{\textbf{\authorTwoName}}\\
		\footnotesize
		\authorTwoAffil
	\end{textblock}	
	\begin{textblock}{\authorCellWidth}[0,0](\colthreeX,\rowoneY)
		\centering
		\small{\textbf{\authorThreeName}}\\
		\footnotesize
		\authorThreeAffil
	\end{textblock}	
}
\defcase{4}{
	\begin{textblock}{\authorCellWidth}[0,0](\coloneX,\rowoneY)
		\centering
		\small{\textbf{\authorOneName}}\\
		\footnotesize
		\authorOneAffil
	\end{textblock}	
	\begin{textblock}{\authorCellWidth}[0,0](\coltwoX,\rowoneY)
		\centering
		\small{\textbf{\authorTwoName}}\\
		\footnotesize
		\authorTwoAffil
	\end{textblock}	
	\begin{textblock}{\authorCellWidth}[0,0](\colthreeX,\rowoneY)
		\centering
		\small{\textbf{\authorThreeName}}\\
		\footnotesize
		\authorThreeAffil
	\end{textblock}
	\begin{textblock}{8}[0,0](\colsinglecentralX,\rowtwoY)
		\centering
		\small{\textbf{\authorFourName}}\\
		\footnotesize
		\authorFourAffil
	\end{textblock}	
}
\defcase{5}{
	\begin{textblock}{\authorCellWidth}[0,0](\coloneX,\rowoneY)
		\centering
		\small{\textbf{\authorOneName}}\\
		\footnotesize
		\authorOneAffil
	\end{textblock}	
	\begin{textblock}{\authorCellWidth}[0,0](\coltwoX,\rowoneY)
		\centering
		\small{\textbf{\authorTwoName}}\\
		\footnotesize
		\authorTwoAffil
	\end{textblock}	
	\begin{textblock}{\authorCellWidth}[0,0](\colthreeX,\rowoneY)
		\centering
		\small{\textbf{\authorThreeName}}\\
		\footnotesize
		\authorThreeAffil
	\end{textblock}
	\begin{textblock}{\authorCellWidth}[0,0](\colDoubleLeftX,\rowtwoY)
		\centering
		\small{\textbf{\authorFourName}}\\
		\footnotesize
		\authorFourAffil
	\end{textblock}	
	\begin{textblock}{\authorCellWidth}[0,0](\colDoubleRightX,\rowtwoY)
		\centering
		\small{\textbf{\authorFiveName}}\\
		\footnotesize
		\authorFiveAffil
	\end{textblock}	
}
\defcase{6}{
	\begin{textblock}{\authorCellWidth}[0,0](\coloneX,\rowoneY)
		\centering
		\small{\textbf{\authorOneName}}\\
		\footnotesize
		\authorOneAffil
	\end{textblock}	
	\begin{textblock}{\authorCellWidth}[0,0](\coltwoX,\rowoneY)
		\centering
		\small{\textbf{\authorTwoName}}\\
		\footnotesize
		\authorTwoAffil
	\end{textblock}	
	\begin{textblock}{\authorCellWidth}[0,0](\colthreeX,\rowoneY)
		\centering
		\small{\textbf{\authorThreeName}}\\
		\footnotesize
		\authorThreeAffil
	\end{textblock}
	\begin{textblock}{\authorCellWidth}[0,0](\coloneX,\rowtwoY)
		\centering
		\small{\textbf{\authorFourName}}\\
		\footnotesize
		\authorFourAffil
	\end{textblock}	
	\begin{textblock}{\authorCellWidth}[0,0](\coltwoX,\rowtwoY)
		\centering
		\small{\textbf{\authorFiveName}}\\
		\footnotesize
		\authorFiveAffil
	\end{textblock}	
	\begin{textblock}{\authorCellWidth}[0,0](\colthreeX,\rowtwoY)
		\centering
		\small{\textbf{\authorSixName}}\\
		\footnotesize
		\authorSixAffil
	\end{textblock}
}
\defcase{7}{
	\begin{textblock}{\authorCellWidth}[0,0](\coloneX,\rowoneY)
		\centering
		\small{\textbf{\authorOneName}}\\
		\footnotesize
		\authorOneAffil
	\end{textblock}	
	\begin{textblock}{\authorCellWidth}[0,0](\coltwoX,\rowoneY)
		\centering
		\small{\textbf{\authorTwoName}}\\
		\footnotesize
		\authorTwoAffil
	\end{textblock}	
	\begin{textblock}{\authorCellWidth}[0,0](\colthreeX,\rowoneY)
		\centering
		\small{\textbf{\authorThreeName}}\\
		\footnotesize
		\authorThreeAffil
	\end{textblock}
	\begin{textblock}{\authorCellWidth}[0,0](\coloneX,\rowtwoY)
		\centering
		\small{\textbf{\authorFourName}}\\
		\footnotesize
		\authorFourAffil
	\end{textblock}	
	\begin{textblock}{\authorCellWidth}[0,0](\coltwoX,\rowtwoY)
		\centering
		\small{\textbf{\authorFiveName}}\\
		\footnotesize
		\authorFiveAffil
	\end{textblock}	
	\begin{textblock}{\authorCellWidth}[0,0](\colthreeX,\rowtwoY)
		\centering
		\small{\textbf{\authorSixName}}\\
		\footnotesize
		\authorSixAffil
	\end{textblock}
	\begin{textblock}{8}[0,0](\colsinglecentralX,\rowthreeY)
		\centering
		\small{\textbf{\authorSevenName}}\\
		\footnotesize
		\authorSevenAffil
	\end{textblock}
}
\defcase{8}{
	\begin{textblock}{\authorCellWidth}[0,0](\coloneX,\rowoneY)
		\centering
		\small{\textbf{\authorOneName}}\\
		\footnotesize
		\authorOneAffil
	\end{textblock}	
	\begin{textblock}{\authorCellWidth}[0,0](\coltwoX,\rowoneY)
		\centering
		\small{\textbf{\authorTwoName}}\\
		\footnotesize
		\authorTwoAffil
	\end{textblock}	
	\begin{textblock}{\authorCellWidth}[0,0](\colthreeX,\rowoneY)
		\centering
		\small{\textbf{\authorThreeName}}\\
		\footnotesize
		\authorThreeAffil
	\end{textblock}
	\begin{textblock}{\authorCellWidth}[0,0](\coloneX,\rowtwoY)
		\centering
		\small{\textbf{\authorFourName}}\\
		\footnotesize
		\authorFourAffil
	\end{textblock}	
	\begin{textblock}{\authorCellWidth}[0,0](\coltwoX,\rowtwoY)
		\centering
		\small{\textbf{\authorFiveName}}\\
		\footnotesize
		\authorFiveAffil
	\end{textblock}	
	\begin{textblock}{\authorCellWidth}[0,0](\colthreeX,\rowtwoY)
		\centering
		\small{\textbf{\authorSixName}}\\
		\footnotesize
		\authorSixAffil
	\end{textblock}
	\begin{textblock}{\authorCellWidth}[0,0](\colDoubleLeftX,\rowthreeY)
		\centering
		\small{\textbf{\authorSevenName}}\\
		\footnotesize
		\authorSevenAffil
	\end{textblock}
	\begin{textblock}{\authorCellWidth}[0,0](\colDoubleRightX,\rowthreeY)
		\centering
		\small{\textbf{\authorEightName}}\\
		\footnotesize
		\authorEightAffil
	\end{textblock}
}
\defcase{9}{
	\begin{textblock}{\authorCellWidth}[0,0](\coloneX,\rowoneY)
		\centering
		\small{\textbf{\authorOneName}}\\
		\footnotesize
		\authorOneAffil
	\end{textblock}	
	\begin{textblock}{\authorCellWidth}[0,0](\coltwoX,\rowoneY)
		\centering
		\small{\textbf{\authorTwoName}}\\
		\footnotesize
		\authorTwoAffil
	\end{textblock}	
	\begin{textblock}{\authorCellWidth}[0,0](\colthreeX,\rowoneY)
		\centering
		\small{\textbf{\authorThreeName}}\\
		\footnotesize
		\authorThreeAffil
	\end{textblock}
	\begin{textblock}{\authorCellWidth}[0,0](\coloneX,\rowtwoY)
		\centering
		\small{\textbf{\authorFourName}}\\
		\footnotesize
		\authorFourAffil
	\end{textblock}	
	\begin{textblock}{\authorCellWidth}[0,0](\coltwoX,\rowtwoY)
		\centering
		\small{\textbf{\authorFiveName}}\\
		\footnotesize
		\authorFiveAffil
	\end{textblock}	
	\begin{textblock}{\authorCellWidth}[0,0](\colthreeX,\rowtwoY)
		\centering
		\small{\textbf{\authorSixName}}\\
		\footnotesize
		\authorSixAffil
	\end{textblock}
	\begin{textblock}{\authorCellWidth}[0,0](\coloneX,\rowthreeY)
		\centering
		\small{\textbf{\authorSevenName}}\\
		\footnotesize
		\authorSevenAffil
	\end{textblock}	
	\begin{textblock}{\authorCellWidth}[0,0](\coltwoX,\rowthreeY)
		\centering
		\small{\textbf{\authorEightName}}\\
		\footnotesize
		\authorEightAffil
	\end{textblock}	
	\begin{textblock}{\authorCellWidth}[0,0](\colthreeX,\rowthreeY)
		\centering
		\small{\textbf{\authorNineName}}\\
		\footnotesize
		\authorNineAffil
	\end{textblock}
}
\defcase{10}{
	\begin{textblock}{\authorCellWidth}[0,0](\coloneX,\rowoneY)
		\centering
		\small{\textbf{\authorOneName}}\\
		\footnotesize
		\authorOneAffil
	\end{textblock}	
	\begin{textblock}{\authorCellWidth}[0,0](\coltwoX,\rowoneY)
		\centering
		\small{\textbf{\authorTwoName}}\\
		\footnotesize
		\authorTwoAffil
	\end{textblock}	
	\begin{textblock}{\authorCellWidth}[0,0](\colthreeX,\rowoneY)
		\centering
		\small{\textbf{\authorThreeName}}\\
		\footnotesize
		\authorThreeAffil
	\end{textblock}
	\begin{textblock}{\authorCellWidth}[0,0](\coloneX,\rowtwoY)
		\centering
		\small{\textbf{\authorFourName}}\\
		\footnotesize
		\authorFourAffil
	\end{textblock}	
	\begin{textblock}{\authorCellWidth}[0,0](\coltwoX,\rowtwoY)
		\centering
		\small{\textbf{\authorFiveName}}\\
		\footnotesize
		\authorFiveAffil
	\end{textblock}	
	\begin{textblock}{\authorCellWidth}[0,0](\colthreeX,\rowtwoY)
		\centering
		\small{\textbf{\authorSixName}}\\
		\footnotesize
		\authorSixAffil
	\end{textblock}
	\begin{textblock}{\authorCellWidth}[0,0](\coloneX,\rowthreeY)
		\centering
		\small{\textbf{\authorSevenName}}\\
		\footnotesize
		\authorSevenAffil
	\end{textblock}	
	\begin{textblock}{\authorCellWidth}[0,0](\coltwoX,\rowthreeY)
		\centering
		\small{\textbf{\authorEightName}}\\
		\footnotesize
		\authorEightAffil
	\end{textblock}	
	\begin{textblock}{\authorCellWidth}[0,0](\colthreeX,\rowthreeY)
		\centering
		\small{\textbf{\authorNineName}}\\
		\footnotesize
		\authorNineAffil
	\end{textblock}
	\begin{textblock}{8}[0,0](\colsinglecentralX,\rowfourY)
		\centering
		\small{\textbf{\authorTenName}}\\
		\footnotesize
		\authorTenAffil
	\end{textblock}
}

\newcommand\authorOneName{Aleksandar Milenkoski}
\newcommand\authorOneAffil{
Institute for Program Structures and Data Organization\\
	Karlsruhe Institute of Technology\\
    Karlsruhe, Germany\\
	\emph{milenkoski@kit.edu}
}
\newcommand\authorTwoName{Alexandru Iosup}
\newcommand\authorTwoAffil{
	Faculty of Engineering, Mathematics and Computer Science \\
	Delft University of Technology\\
	Delft, Netherlands\\
	\emph{A.Iosup@tudelft.nl}
}

\newcommand\authorThreeName{Samuel Kounev}
\newcommand\authorThreeAffil{
    Institute for Program Structures and Data Organization\\
	Karlsruhe Institute of Technology\\
    Karlsruhe, Germany\\
	\emph{kounev@kit.edu}
}
\newcommand\authorFourName{Kai Sachs}
\newcommand\authorFourAffil{
	SAP AG\\
	Walldorf, Germany\\
	\emph{kai.sachs@sap.com}
}
\newcommand\authorFiveName{Piotr Rygielski}
\newcommand\authorFiveAffil{
    Institute for Program Structures and Data Organization\\
	Karlsruhe Institute of Technology\\
    Karlsruhe, Germany\\
	\emph{piotr.rygielski@kit.edu}
}
\newcommand\authorSixName{Jason Ding}
\newcommand\authorSixAffil{
	Performance Engineering - Cloud Apps\\
	Salesforce.com\\
	San Francisco, California, USA\\
	\emph{jding@salesforce.com}
}
\newcommand\authorSevenName{Walfredo Cirne}
\newcommand\authorSevenAffil{
	Google Inc.\\
	Mountain View, California, USA\\
	\emph{walfredo@google.com}
}
\newcommand\authorEightName{Florian Rosenberg}
\newcommand\authorEightAffil{
	IBM T.J. Watson Research Center\\
	Skyline Drive Hawthorne, New York, USA\\
	\emph{rosenberg@us.ibm.com}
}
\newcommand\authorNineName{FirstName9 LastName9}
\newcommand\authorNineAffil{
	Department of Cloud Computing3,\\
	State University of SomeCity3,\\
	SomeCity'sLongerName, LongLongLongCountryName4,\\
	e-mail2@e-mail.com
}
\newcommand\authorTenName{FirstName10 LastName10}
\newcommand\authorTenAffil{
	Department of Cloud Computing3,\\
	State University of SomeCity3,\\
	SomeCity'sLongerName, LongLongLongCountryName4,\\
	e-mail2@e-mail.com
}

\begin{document}
 
\selectlanguage{english} 
\frontmatter

\thispagestyle{empty}
\newcommand{\changefont}[3]{\fontfamily{#1} \fontseries{#2} \fontshape{#3} \selectfont}
\newcommand{\diameter}{20}
\newcommand{\xone}{-25}
\newcommand{\xtwo}{165}
\newcommand{\yone}{20}
\newcommand{\ytwo}{-253}

\newcommand{\rowoneY}{5.5}		
\newcommand{\rowtwoY}{7.0}
\newcommand{\rowthreeY}{8.5}
\newcommand{\rowfourY}{10.1}

\newcommand{\coloneX}{2.5}
\newcommand{\coltwoX}{7.45}
\newcommand{\colthreeX}{12.4}

\newcommand{\colDoubleLeftX}{5}
\newcommand{\colDoubleRightX}{10}

\newcommand{\colsinglecentralX}{5.9}

\newcommand{\authorCellWidth}{4.9}

\begin{titlepage}
\begin{tikzpicture}[overlay]
\draw[color=gray]  
 (\xone mm, \yone mm) -- (\xtwo mm, \yone mm) arc (90:0:\diameter pt) 
  -- (\xtwo mm + \diameter pt , \ytwo mm) -- (\xone mm + \diameter pt , \ytwo mm) 
 arc (270:180:\diameter pt) -- (\xone mm, \yone mm);
\end{tikzpicture}

\changefont{phv}{m}{n}	
\begin{textblock}{14}[0,0](3,2.3)
	\centering
	\large{\TRnumber}\\
	\vspace*{1cm}
	\huge{\mytitle}\\
	\vspace*{0.5cm}
	\Large{\WGname}
\end{textblock}
\begin{textblock}{15.5}[0,0](2,5.2)
	\begin{tikzpicture}
		\fill[red!80!brown] (0,0cm) rectangle (19.5cm,0.1cm);
	\end{tikzpicture}
\end{textblock}

\begin{center}
	\putAuthors{\numAuthors}
\end{center}

\begin{textblock}{14}[0,0](3,13)
	\hfill
	\includegraphics[width=3cm]{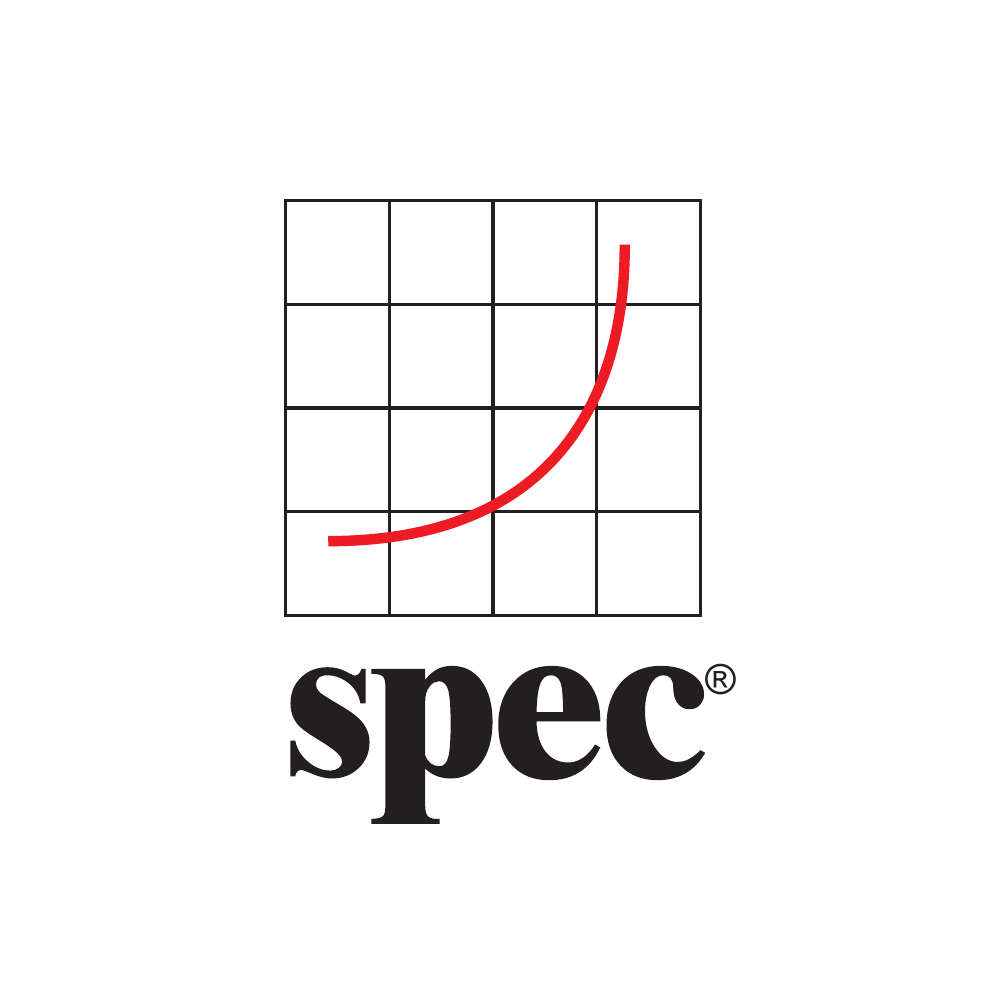} \hfill
	\includegraphics[width=1.9cm]{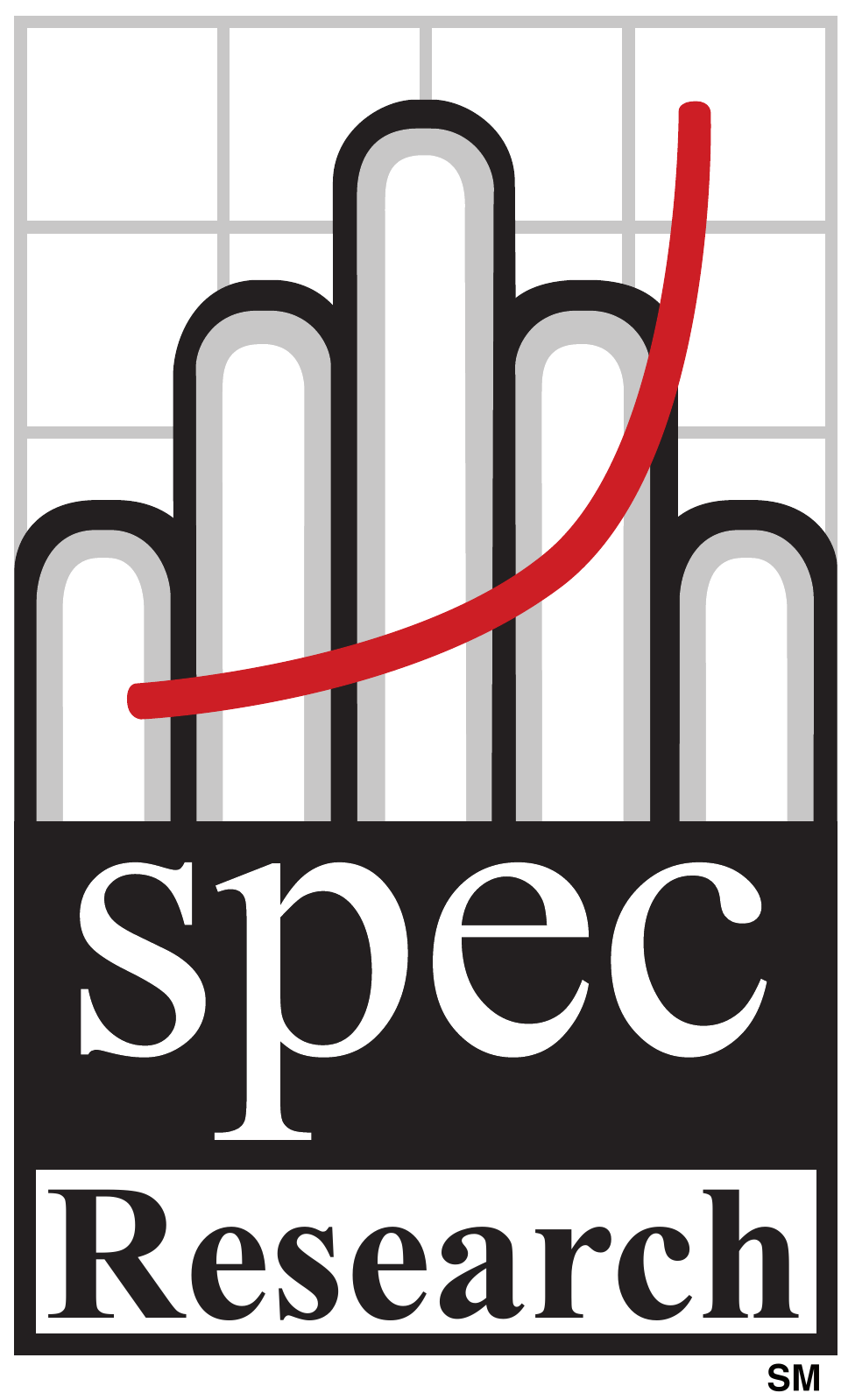} \hspace{1.1cm}\hfill
	\includegraphics[width=1.9cm]{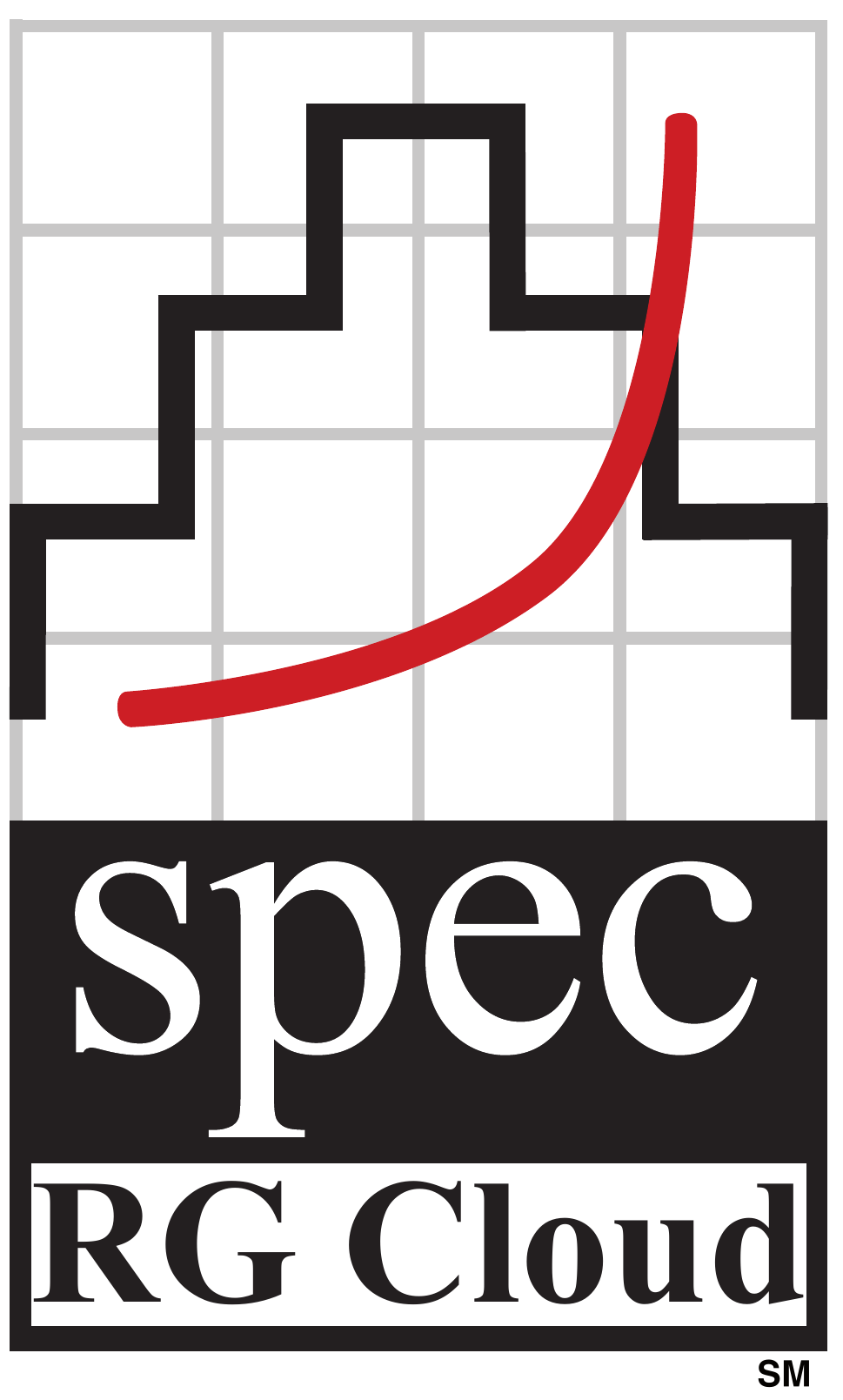} \hspace{1.1cm}\hfill
	\hfill
\end{textblock}

\begin{textblock}{14}[0,0](3,15)
	\noindent\textbf{Acknowledgements}\hfill\vspace{0.5em}\hrule
	\vspace{0.5em}\noindent\footnotesize
	\Acknowledgements
\end{textblock}

\begin{textblock}{14}[0,0](3,16.75)
	\centering
	\large{\textbf{\TRdate}}
	\hfill
	\large{\textbf{\TRcentralURL}}
	\hfill
	\large{\textbf{\TRrightURL}}
\end{textblock}

\end{titlepage}

\newpage 
\thispagestyle{empty}
\mbox{}

\newpage
\pagenumbering{roman}
\setcounter{tocdepth}{4}
\begin{spacing}{1.3}
\tableofcontents
\end{spacing}

\newpage
\thispagestyle{plain}
\section*{Executive Summary}

Cloud computing is becoming an increasingly lucrative branch of the existing information and communication technologies (ICT). Enabling a debate about cloud usage scenarios can help with attracting new customers, sharing best-practices, and designing new cloud services. In contrast to previous approaches, which have attempted mainly to formalize the common service delivery models (i.e., Infrastructure-as-a-Service, Platform-as-a-Service, and Software-as-a-Service), in this work, we propose a formalism for describing common cloud usage scenarios referred to as \emph{cloud usage patterns}. Our formalism takes a structuralist approach allowing decomposition of a cloud usage scenario into elements corresponding to the common cloud service delivery models. Furthermore, our formalism considers several cloud usage patterns that have recently emerged, such as hybrid services and value chains in which mediators are involved, also referred to as value chains with mediators. We propose a simple yet expressive textual and visual language for our formalism, and we show how it can be used in practice for describing a variety of real-world cloud usage scenarios. The scenarios for which we demonstrate our formalism include
resource provisioning of global providers of infrastructure and/or platform resources, online social networking services, user-data processing services, online customer and ticketing services, online asset management and banking applications, CRM (Customer Relationship Management) applications, and online social gaming applications. 

\vspace{0.5cm}

\textbf{Keywords\footnote{The used keywords are defined as part of The 2012 ACM Computing Classification System \cite{acm:classification}.}:} \\
cloud computing; usage patterns; domain-specific language; visual language; structuralism; \\
CCS -  Software and its engineering -  Software notations and tools -  Context specific languages -  Domain specific languages\\
CCS -  Theory of computation -  Formal languages and automata theory -  Grammars and context-free languages
CCS -  Software and its engineering -  Software notations and tools -  Context specific languages -  Visual languages\\
CCS -  Computer systems organization -  Architectures -  Distributed architectures -  Cloud computing\\
CCS -  Applied computing -  Enterprise computing -  Service-oriented architectures\\
CCS -  Applied computing -  Enterprise computing -  Business process management -  Cross-organizational business processes\\
CCS -  Software and its engineering -  Software creation and management -  Software development process management -  Software development methods -  Design patterns\\

\mainmatter
\include{intro}

\newpage
\pagenumbering{arabic}
\setcounter{page}{1}

\section{Introduction}\label{sec:intro}

The cloud computing paradigm is constantly gaining in popularity, mainly due to the reported
numerous benefits for cloud users such as the ease-of-use, the on-demand resource provisioning, the pay-per-use business model, and the ability of cloud environments to support execution of applications of various types \cite{gens:cloud}. Recent in-depth studies (e.g., Broderick \cite{broderick:worldwide}) predict rapid expansion of the cloud computing market in the years to come in many areas where computing capabilities are needed. The adoption of clouds in many application areas, such as online collaborative and entertainment services, and business intelligence and data preservation, led to broadening of the term ``cloud''. At the time of writing, i.e., 2013, there are more than 15 different definitions of broad acceptance~\cite[p.36]{book/KeuperOD11}; unsurprisingly, this leads to confusion of terms and hampers discussions about cloud usage for all stakeholders of the cloud ecosystem, such as users, system integrators, and service brokers. The cloud ecosystem includes many stakeholders - participants which consume and/or deliver resource provisioning services, and which generate or transfer value. For instance, Leimeister et al. \cite{leimeister+2010} identify the following stakeholders: customers (i.e., buyers of services); service providers (i.e., developers and operators of services that offer value to customers); infrastructure providers (i.e., providers of technical backbones for generation or transfer of value to customers); consultants (i.e., providers of customer support for selection and implementation of services operated by service providers), and so on. To contribute towards addressing the previously mentioned issue, we propose and investigate a formalism for describing cloud usage scenarios in which stakeholders of the cloud ecosystem are involved. In the context of this work, we refer to the proposed formalism as \emph{cloud usage patterns}.

The formalism that we propose allows to describe in an agreed-upon notation the abstract and the practical usage of cloud resource provisioning services in a given cloud usage scenario for any application or application domain. For instance, the use of leased infrastructure resources by Go!Animate \cite{GoAnimate} (e.g., computing, storage resources) from the infrastructure provider Amazon (or, to be specific, Amazon Web Services \cite{AmazonWebServices}) for servicing users processing amateur videos, can be formally described with the string \texttt{i$_3$.s.e$_{6}$}, where each letter relates to one of the known cloud service delivery models \cite{mell+2010} (i.e., Infrastructure-as-a-Service and Software-as-a-Service in the given string), and/or a stakeholder (i.e., an end-user). Also, each number relates to a volume of provisioned resources that may be expressed, for example, as a logarithmic value (e.g., 3 for 1,000 in the given string). The formalism that we propose provides many significant advantages for the stakeholders of the cloud ecosystem. For instance, it greatly simplifies the discussion about generic and specific cloud usage scenarios, which, among others, affects service providers, customers, and consultants. Further, potential and actual users of cloud environments can use our formalism to specify their service requirements in a compact manner. Also, cloud system designers can easily compile frequently used patterns, whereas cloud integrators and auditors can benefit by sharing cloud usage best-practices from industry. In addition, service providers can learn how to tune systems, researchers and consultants can classify and compare different scenarios, engineers can be trained for the most common usage patterns of cloud environments, and so on.

Other fields have already made significant progress towards standardization of their common use cases and have extensively used scenario formalisms. One of the earliest formalisms with a long-lasting impact is Shannon's theory of communication, which introduced the main components of communication and proposed abstract textual and graphical notations for many common processes in communication. In architectural design, Alexander~\cite{book/Alexander64} abstracted tens of patterns common in the field. In creative industries, the structuralist formalism of folk tales by Propp~\cite{book/Propp68}, later extended by Campbell~\cite{book/Campbell72} and Vogler~\cite{book/Vogler07}, has been used in scriptwriting at Hollywood \cite{book/McKee99}. In software engineering, the use of patterns and also anti-patterns has become wide-spread, following the seminal work of ``the gang of four''~\cite{book/GammaHJV94}. Each of the previously mentioned formalisms has acted as a catalyst for a fragmented market or practice. 

Our main objective in this work is the design of a formalism for describing cloud usage scenarios and also the demonstration of its practicality by applying it to a comprehensive set of such scenarios. Towards this goal, our main contribution is an abstract formalism that uses textual and visual notation, interchangeably, to express succinctly yet understandably various such scenarios. Our formalism uses less than ten symbols to specify involved stakeholders and their roles, service delivery models, service level agreements (SLAs), and sizes/volumes of provisioned resources, as part of a cloud usage pattern to which multiple cloud usage scenarios can conform. We designed our formalism considering several design requirements such as expresiveness, mutual exclusiveness, and comprehensibility, aiming to satisfy all of these requirements while taking into account the trade-offs between them (e.g., the trade-off between expresiveness and comprehensibility). We show evidence that our formalism can be used in practice for describing real-world cloud usage scenarios; that is, we use our formalism to describe the use of an airline company's booking system deployed in a cloud environment, of an asset management system, of a CRM software delivery service, and so on. We also analyze relationships between cloud applications and application types involved in real-world cloud usage scenarios, on the one hand, and the respective cloud usage patterns, on the other hand, in order to identify and analyze best practices in industry.

The remainder of this document is structured as follows: 
In Section~\ref{sec:apps}, we categorize applications deployed in cloud environments and we also make a selection of several representative applications; in Section~\ref{sec:cloud-usage-patterns}, we analyze the main dimensions across which we discuss cloud usage scenarios and propose a formalism for describing such scenarios addressing all of the considered dimensions; in Section~\ref{sec:practice}, we use the formalism to describe several real-world cloud usage scenarios, many of which involve usage of the applications that we selected in Section~\ref{sec:apps}; in Section~\ref{sec:related}, we put our contribution in the context of related work, and finally, in Section~\ref{sec:conclusions}, we conclude our work. 

\subsection{Terms and Definitions}
\label{sec:terms-and-definitions}

In the context of this work, a concrete and accurate definition of a cloud system and of the different services that such a system may offer is needed. Such definitions are crucial for building a formalism describing cloud usage scenarios, which requires a stricly defined understanding of the features and characteristics of cloud systems, and in addition, of the different cloud service delivery models. To this end, we take into consideration the definitions of the term ``cloud system'' and of the common service delivery models as proposed by Mell et al. \cite{mell+2010} from NIST (National Instutite for Standards and Technology). Next, we briefly present these definitions.

\textbf{Cloud system and infrastructure.} Mell et al. \cite{mell+2010} define a cloud system as a collection of hardware and software enabling the five essential characteristics of cloud computing - on-demand self-service, broad network access, resource pooling, rapid elasticity, and measured service. We refer the reader to \cite{mell+2010}, or to the Glossary of this work, for detailed information on these characteristics. Further, Mell et al. \cite{mell+2010} define a cloud infrastructure as an infrastructure consisting of a physical and an abstract layer. The physical layer is consisting of the bare hardware resources, while the abstract layer is a specialized software that enables the previously mentioned essential characteristics of cloud computing. 

\textbf{Cloud service delivery models.}  Mell et al. \cite{mell+2010} differentiate between three different service delivery models, i.e., Infrastructure-as-a-Service (IaaS), Platform-as-a-Service (PaaS), and Software-as-a-Service (SaaS), according to which infrastructure, platform, and software resources are provisioned, respectively\footnote{We refer the reader to the Glossary at the end of the document for definition of infrastructure, platform, and software resources.}. In the following, we give a brief description of each of these models:

\begin{quote}
(i) Software-as-a-Service (SaaS): This model enables provisioning of applications deployed in cloud environments to end-users. An end-user is not able to manage and/or control infrastructure resources (e.g., servers, network, storage resources), nor the environment in which the provisioned application is hosted. 
\end{quote}

\begin{quote}
(ii) Platform-as-a-Service (PaaS): This model enables an end-user to deploy customer-created or aquired applications created using libraries, services, languages, tools, and similar, that may, but do not have to, be provided by the cloud provider itself that enables this model. An end-user is not able to manage and/or control infrastructure resources (e.g., servers, network, storage), but can control application-hosting environments.
\end{quote}

\begin{quote}
(iii) Infrastructure-as-a-Service (IaaS): This model enables provisioning of basic infrastructure resources (e.g., servers, network, storage) such that an end-user may deploy arbitrary software, including operating systems. An end-user is not able to manage and/or control the underlying cloud infrastructure.
\end{quote}
\newpage 

\section{A Representative Set of Cloud Applications}
\label{sec:typical-apps-on-clouds}
\label{sec:apps}

In this section, we provide a categorization of applications deployed in cloud environments (i.e., cloud applications), and then, we reason about a selection of representative cloud applications, which are used throughout the rest of this  work as use cases. In Section~\ref{sec:apps:taxonomy}, we categorize and discuss a variety of cloud applications, and in Section~\ref{sec:apps:selection}, we select several such applications.

\subsection{A Categorization of Cloud Applications}\label{sec:apps:taxonomy}

The categorization we propose in the following includes both traditional applications that have been migrated to clouds as well as applications that have been developed exclusively for cloud environments. The proposed categorization does not include applications that have not yet been migrated to clouds, such as applications tightly bound to hardware (e.g., hardware monitoring and control software) and low-latency applications (e.g., racing and sport games, first-person simulations normally used in military training). Our contribution is to show evidence of the diversity of cloud-hosted applications, thus enabling a judicious selection of use cases.

For categorizing cloud applications, we use the taxonomy of software applications proposed by Forward and Lethbridge~\cite{Forward2008}\footnote{The Forward and Lethbridge taxonomy \cite{Forward2008} is described in more detail in Section~\ref{sec:related:apps}.}; for a critical analysis of alternative taxonomies, such as the ACM computing taxonomy and GoogleCode's application tags, we refer the reader to Forward and Lethbridge~\cite[Section 1.2]{Forward2008}. The taxonomy of Forward and Lethbridge \cite{Forward2008} includes nearly 200 application types grouped into four major categories: data-dominant software (e.g., consumer-oriented software, business-oriented software, design and engineering software), systems software, control-dominant software, and computation-dominant software (e.g., operations research and scientific software). The data-dominant and computation-dominant categories (depicted as categories \textbf{A} and \textbf{D} in Figure~\ref{fig:general-soft-taxonomy}, Section~\ref{sec:related:apps}) include many applications that are normally seen in cloud environments. Given that systems software and control-dominant applications (categories \textbf{B} and \textbf{C} in the taxonomy of Forward and Lethbridge~\cite{Forward2008}) are normally used in relation with specific hardware, we do not foresee them as immediate candidates for deployment in cloud environments, with the exception of simple enablers of larger applications (e.g., load balancers, messaging queues, and so on). 

The categorization that we propose, depicted in Table~\ref{tab:app-types-in-cloud}, shows that many cloud applications fit well into the taxonomy proposed by Forward and Lethbridge~\cite{Forward2008}. However, there are cases in which a cloud application can only be described as a combination of several categories from the Forward and Lethbridge taxonomy \cite{Forward2008}. For example, the data mining services packaged as the integrated cloud service Google BigQuery or Cloud9 Analytics, and the database services packaged as Amazon DB and Quickbase, can fit well in the following categories: spreadsheets/calculators (category A.con.2.c); personal management (A.con.4, with several matching sub-categories); strategic and operations analysis (A.bus.1, with several matching sub-categories); data warehousing (A.bus.3.a); and data mining and business intelligence (A.bus.3.h). We argue that such an ambiguous matching is due to the high complexity of modern cloud applications, which in turn is a result of the growing sophistication of users and application domains. Software providers in their efforts to attract new customers and to retain existing customers, often extend the main functionality of an application with features that may be categorized differently from the original goals of the application. Moreover, bundling services that combine several services into a single offering are typical for cloud environments. We discuss value chains in the context of cloud computing and service bundling  in more detail in Section~\ref{sec:cloud-usage-patterns}.

\begin{longtable}{|P{3cm}|P{8cm}|P{4cm}|}
\caption{categorization of cloud applications (the listed categories are matched with respective categories in the Forward and Lethbridge taxonomy~\cite{Forward2008} enclosed in square brackets. Application types are sorted ascendingly according to the matching category in the Forward and Lethbridge taxonomy.)}
 \label{tab:app-types-in-cloud}\\
\hline \multicolumn{1}{|p{3cm}|}{\textbf{Application Type}} & \multicolumn{1}{p{8cm}|}{\textbf{Usage Description}} & \multicolumn{1}{p{3cm}|}{\textbf{Example}} \\ \hline 
\endfirsthead

\multicolumn{3}{c}%
{{\tablename\ \thetable{} -- continued from previous page}} \\
\hline \multicolumn{1}{|p{3cm}|}{\textbf{Application Type}} & \multicolumn{1}{p{8cm}|}{\textbf{Usage Description}} & \multicolumn{1}{p{3cm}|}{\textbf{Example}} \\ \hline 
\endhead

\hline \multicolumn{3}{|r|}{{\small Continued on next page}} \\ \hline
\endfoot

\hline
\endlastfoot
Messaging [A.con.1.a--c] & Enable users to exchange information normally in text form and of relatively small volume.
& Various e-mail and instant messengers\\ 
\hline
File storage and exchange [A.con.1.f] & Enable users to store and exchange files. 
These applications are normally deployed with sophisticated security and privacy policies.
They usually require substantial amount of storage and network capabilities. & Dropbox, RapidShare, Apple iCloud, Google Drive, Microsoft SkyDrive\\
\hline
Data mining [A.con.2.c, A.con.4, A.bus.3.a/h] & Analyse a dataset to discover relations between data. These applications normally analyze large amounts of data in order to find non-trivial patterns and relations. They usually generate CPU- and file I/O-intensive workloads. 
 & Cloud9 Analytics, Aster Database, Google BigQuery\\
\hline
Databases [A.con.2.c, A.con.4, A.bus.3.a/h] & Store data in a structured and organized manner. Database applications feature data storage and management mechanisms, which require a significant amount of storage and processing capabilities. & Amazon SimpleDB, Quickbase \\
\hline
Massive Open Online Course (MOOC) [A.con.3.b, etc.] & Enable students to access lecture material, such as (video) presentations and assignments, work in online study groups (forums), grade each other's assignments and receive electronic feedback, authenticate and take exams, and so on. MOOCs are Web-based online courses designed for large-scale enrollment, typically at university level, but without accreditation or credit-offering~\cite{mooc:nytimes12}.  MOOC platforms might also offer additional services for assignment processing, automatic cheat detection, and automatic grading.
& edX, Coursera, Udacity\\
\hline
User data processing [A.con.3.d, etc.] & Enable users to work on user-supplied files by providing professional tools for various tasks, e.g., video processing, audio processing, and document creation. 
& Go!Animate, Animoto, Flickr\\
\hline
Audio/Video Streaming [A.con.3.e] & Enable real-time exchange of multimedia data. The applications that belong to this category normally require fast network connectivity for efficiency. Further, if a multimedia stream is processed during streaming, these applications also require a substantial amount of processing capabilities.  & VideoOnDemand, Netflix, Spotify, SoundCloud\\
\hline
Online gaming and meta-gaming [A.con.3.g, A.inf.3.a/c/d] & Provide entertainment for a group of users sharing the same virtual reality. Many games have social elements; that is, they enable social mechanisms such as friendship and foster pro-social gaming emotions (e.g., competitiveness, vicarious pride, and similar). The meta-gaming element allows users to share information, opinions, images, and videos related to the games; in this sense these applications are similar to messaging and collaboration applications. Some applications may provide only video feedback, as opposed to the traditional stream of server-initiated binary commands; these services are similar to audio/video streaming applications, but are focused on games and may include specific adaptation to some audio/video channels, e.g., higher quality for the voice commands issued by a team-leader to the other players. & World of Warcraft, OGame, Zynga, GoLive, Gaikai\\
\hline
Financials [A.bus.1.b/f, A.bus.2.c/d, A.des.3.p, etc.] & Manage expenses. The applications that belong to this category normally use a back-end database and a business analytics engine. The security of the stored and transmitted data during operation is crucial due to high sensitivity. 
& Workday, Expensify\\
\hline
Customer Relationship Management / Partner Relationship Management (CRM/PRM) [A.bus.1.f, A.bus.1.e/f, etc.] & Manage contacts with business partners and customers. The applications of this type are focusing on data integration and storage. Similar to financial applications, the stored data by CRM/PRM applications normally contains sensitive information such as employees' and/or customers' personal data. & Salesforce.com, NetSuite, Basecamp\\
\hline
Project and team management [A.bus.1.g] & Allow users to manage teams and projects by defining roles, formulating and scheduling tasks, packaging and managing products, and similar. & Bubbl.us, Huddle, LiquidPlanner, Zoho Planner, Teambox, Agilezen\\
\hline
On-line Transaction Processing (OLTP) [A.bus.3.c, A.bus.4.a--j] & Process user-supplied datasets by executing high numbers of simple instructions. The data being processed is usually related to business-oriented tasks running in a cloud environment. & Amazon RDS, CloudTran OLTP Sandbox, Apache Pig\\
\hline
Collaboration [A.des.1.b] & Enable users to communicate and collaboratively read and/or write to a single document or a multimedia file. 
& WebEx, Google Docs, Blogspot, Wordpress, Stormboard, Wallwisher, Adobe Creative Cloud\\
\hline
Development environments [A.des.1--11] & Provide a development environment that may also support collaboration between multiple users. The applications of this type enable code editing, automatic code analysis, product testing, bug tracking and removal, and so on. 
& Cloud9, Ajax.org's Ace, Codeanywhere, Kodingen, CodeRun Studio\\
\hline
Content management [A.inf.3.c/k] & Enable web content management, interactivity, and navigation. The applications of this type support creation, design, and maintenance of web pages. 
 & Clickability, Crown Point Design\\
\hline
Social networking [A.inf.3.d] & Enable users to exchange messages and multimedia as members of a virtual society normally accessed through a web front-end. 
& Facebook, Reddit\\
\hline
Online commerce [A.inf.3.o] & Enable product presentation and sale. The applications of this type are similar to social networking applications (i.e., users as members of a virtual society present, describe, and comment on sale items), however, the goal is to sell and/or buy products so the non-functional requirements put on the application's operation are different. Many of these applications feature sophisticated product search and comparison mechanisms.
 & Amazon, eBay, marktplaats.nl\\
\hline
High Performance Computing (HPC) [D.or.2, D.im, D.sci] & Offer services to solve complex computational problems, usually with application to scientific and engineering studies. These services may consist of libraries or even integrated applications. The workloads of the applications belonging to this category are normally characterized as CPU- and memory- intensive. & WolframAlpha\\
\hline
e-Science [D.or.2, D.sci] & Provide computational services for scientific processes: control of scientific instruments, production of data from simulations, gathering and reducing data, analyzing and modeling results, and visualizing results. These services include provisioning of CPU cycles and disk storage, value-adding services (e.g., checkpointing and backup), complex services (e.g., scientific workflow management), and so on. 
& Netherlands eScience Center, NERSC, NorduGRID, Open Science Grid\\
\hline
Integration [B.mid] & Manage cooperation between multiple applications and services supporting composition of new software products. 
& Boomi AtomSphere, Gnip\\
\hline
Web hosting services [B.svr]  & Provide web hosting functionalities. The applications in this category feature provisioning of servers for web hosting, applications servers, and/or infrastructure resources (e.g., storage resources).
 & Eleven2, OVH Cloud\\
\hline
\end{longtable}

\subsection{Application Selection} \label{sec:apps:selected}\label{sec:apps:selection}

Throughout this work, we consider a set of selected software applications provisioned to end-users according to the SaaS service delivery model, as use cases; we use these applications in the context of relevant resource provisioning scenarios in cloud environments (i.e., cloud usage scenarios) when analyzing and identifying the benefits of different approaches for software provisioning (i.e., application provisioning scenarios) in terms of the involved service delivery models, cloud providers, SLAs, and similar. We use the selected applications to analyze relationships between application provisioning scenarios in cloud environments, on the one hand, and the respective cloud usage patterns to which these scenarios conform, on the other hand. We provide such an analysis in Section~\ref{sec:analysis-patters-applications}.

We show the selected applications and the respective application types in Table~\ref{tab:apps:selected}. The application types listed in Table~\ref{tab:apps:selected} are defined as part of the application categorization that we provided in Section~\ref{sec:apps:taxonomy}. 

\begin{table}[ht]
\caption{the selected applications}
\centering
\begin{tabular}{|p{3.5cm}|p{6cm}|p{8cm}|}
\hline
Application & Application Type \\
\hline \hline
Facebook \cite{Facebook} & Social networking \\ \hline
Go!Animate \cite{GoAnimate} & User data processing \\ \hline
EasyJet \cite{easyJet} & CRM/PRM \\ \hline
Salesforce.com \cite{SalesForce} &  CRM/PRM  \\ \hline
Zynga \cite{Zynga} & Online gaming and meta-gaming \\ 
\hline
\end{tabular}
\label{tab:apps:selected}
\end{table} 

We selected the applications listed in Table~\ref{tab:apps:selected} according to the following relevant criteria:

(i) Representative application types: The selected applications are of types that are commonly seen in cloud environments. For instance, the IBM Institute for Business Value \cite{ibm:dispelling} has performed an extensive survey of applications deployed in cloud environments identifying the application types listed in Table~\ref{tab:apps:selected} as amongst the most common ones. 

(ii) Representative workload characteristics: Under representative workload characteristics, we understand characteristics that exercise the unique features of cloud environments, such as scalability and on-demand resource provisioning. Some of these characteristics include workload transiency, sudden peaks in workload intensity, and similar. Given the large user-base of the applications listed in Table~\ref{tab:apps:selected} (e.g., Facebook \cite{Facebook}, Salesforce.com \cite{SalesForce}), and the fact that applications of the same types are commonly seen in cloud environments \cite{ibm:dispelling}, one may conclude that the workloads of the selected applications have characteristics that exercise the unique features of cloud environments (e.g., intensity peaks due to large number of concurrent users, and so on). 

(iii) Publicly available records supporting the analysis of application provisioning scenarios in cloud environments: 
This includes relevant press release statements, technical reports, white papers, and similar. We support our analysis by referring to such records in Section~\ref{sec:analysis-patters-applications}, where we discuss various application provisioning scenarios.

\newpage 

\section{Cloud Usage Patterns: A Formalism for Description of Cloud Usage Scenarios}
\label{sec:cloud-usage-patterns}
\FloatBarrier

In this section, we propose a formal notation language for describing common real-world cloud usage scenarios. A specification of a cloud usage scenario by means of our proposed formalism is referred to as \emph{cloud usage pattern}. The proposed cloud usage patterns can be used to present and describe real-world cloud usage scenarios in a formal and standard manner. They are able to specify both simple cloud usage scenarios (i.e., where a single service provider provisions resources directly to end-users), as well as more complex scenarios (i.e., where multiple service providers, belonging to a single or multiple legal entities, collaborate and mutually interact in order to provision resources to an end-user). From a business perspective, the latter are known as value chains, or value networks, since multiple stakeholders are involved each of them adding value to end-user \cite{leimeister+2010}. Quoting from Leimeister et al. \cite{leimeister+2010}: ``\emph{... network of relationships that generates economical value and other advantages through complex dynamical exchanges between companies.}'' Standardized and formal descriptions of cloud usage scenarios can be exploited in many areas such as cloud performance benchmarking where, for example, the specification of formally defined and structured information about a SUT (System Under Test, i.e., in the context of this work - an evaluated cloud environment) is typically required. For instance, relevant information about a SUT is information on the interaction between the different service delivery models (i.e., IaaS, PaaS, and SaaS) involved in a service provisioning scenario. The complexity of mandating structured and standardized relevant information about the service delivery models of cloud environments evaluated as part of benchmarking efforts is acknowledged as an important issue in a recent report of SPEC's OSG Cloud Subcommittee \cite{OSGCloud12}. 

The cloud usage patterns we propose convey information on several aspects relevant for the formal description of cloud usage scenarios, referred to as \emph{dimensions}. In the following, we first present the different dimensions and discuss them in detail. Then, we present our formalism for describing cloud usage scenarios, which can take a textual or visual form. We propose cloud usage patterns enabling formal specification of simple cloud usage scenarios, referred to as \emph{elementary} cloud usage patterns (Section~\ref{sec:elementarycups}), and of more complex cloud usage scenarios, including specification of hybrid resource provisioning (Section~\ref{sec:cupdsforhybrid}) and of value chains with mediators (Section~\ref{sec:cupsforvaluechains}), referred to as \emph{extended} cloud usage patterns. 

\subsection{Dimensions of Cloud Usage Patterns}
\label{sec:dimensions:cups}

In this section, we identify and categorize relevant information for the formal description of cloud usage scenarios. The proposed cloud usage patterns are designed to describe all common scenarios in which a cloud infrastructure may be used. As an example, the provisioning of infrastructure, platform, and/or software resources is managed with respect to many Quality-of-Service (QoS) requirements and customer contracts defined as part of Service-Level Agreements (SLAs). The latter may have a significant impact on the behavior of the infrastructure, platform, or software resource provisioning mechanisms of cloud environments. 
Also, an end-user for executing a given task may use services from only a single IaaS, PaaS, or SaaS provider, or might use resources from multiple such providers at the same time which may, or may not be, within the same legal boundary. For instance, Staten et al. \cite{Staten2011} state that the once clearly distinguishable IaaS, PaaS, and SaaS service delivery models have recently begun to overlap forming so-called \emph{mixed service delivery models}, which are gaining on popularity among cloud service providers. From a business perspective, the mixed service delivery models are referred to as value chains, previously mentioned in Section~\ref{sec:cloud-usage-patterns}. An example of a cloud provider featuring a mixed service delivery model is Microsoft Azure, a former pure PaaS provider that is now known to support provisioning of platform resources with underlying IaaS services to enhance efficiency and add value to end-users (see \cite{windowsazure:infrastructure}). Driven by the increasing customer demand for greater flexibility, the number of cloud providers offering mixed service delivery models are expected to increase in the years to come. 

In light of the above discussion, one may conclude that when describing a cloud usage scenario,  information on various aspects, such as the relationships (e.g., SLAs) between the different stakeholders (e.g., end-users, cloud providers) and service delivery models (i.e., IaaS, PaaS, and SaaS) involved in executing end-user tasks, is required. In Figure~\ref{fig:DimensionsCategories}, we depict the dimensions and their categories that we use as a basis when constructing cloud usage patterns. Under dimension, we understand a relevant information on a specific aspect relevant for description of real-world cloud usage scenario.

\begin{figure}[ht]
\begin{minipage}{\textwidth}
\fontsize{11pt}{11pt}\selectfont 
\textbf{Stakeholders} \\
--- \textbf{End-user} \\
--- \textbf{Organization} \\
------ \textbf{Cloud service provider} \\
--------- Native provider\\
--------- Non-native provider\\
------ \textbf{End-user's organization} \\

\textbf{Abstraction Levels} \\
--- \textbf{SaaS} \\
--- \textbf{PaaS} \\
--- \textbf{IaaS} \\
--- \textbf{Hardware Resources}\\
------ With virtualization technology \\
------ Without virtualization technology \\

\textbf{Roles} \\
--- Consumer \\
--- Provider \\
--- Intermediary \\

\textbf{SLAs} \\
--- Internal \\
--- External \\

\textbf{Size/Volume} \\
\caption{dimensions and respective categories of a cloud usage pattern}
\label{fig:DimensionsCategories}
\end{minipage}
\end{figure}
\begin{figure}[ht]
\begin{center}
\includegraphics[scale=0.6]{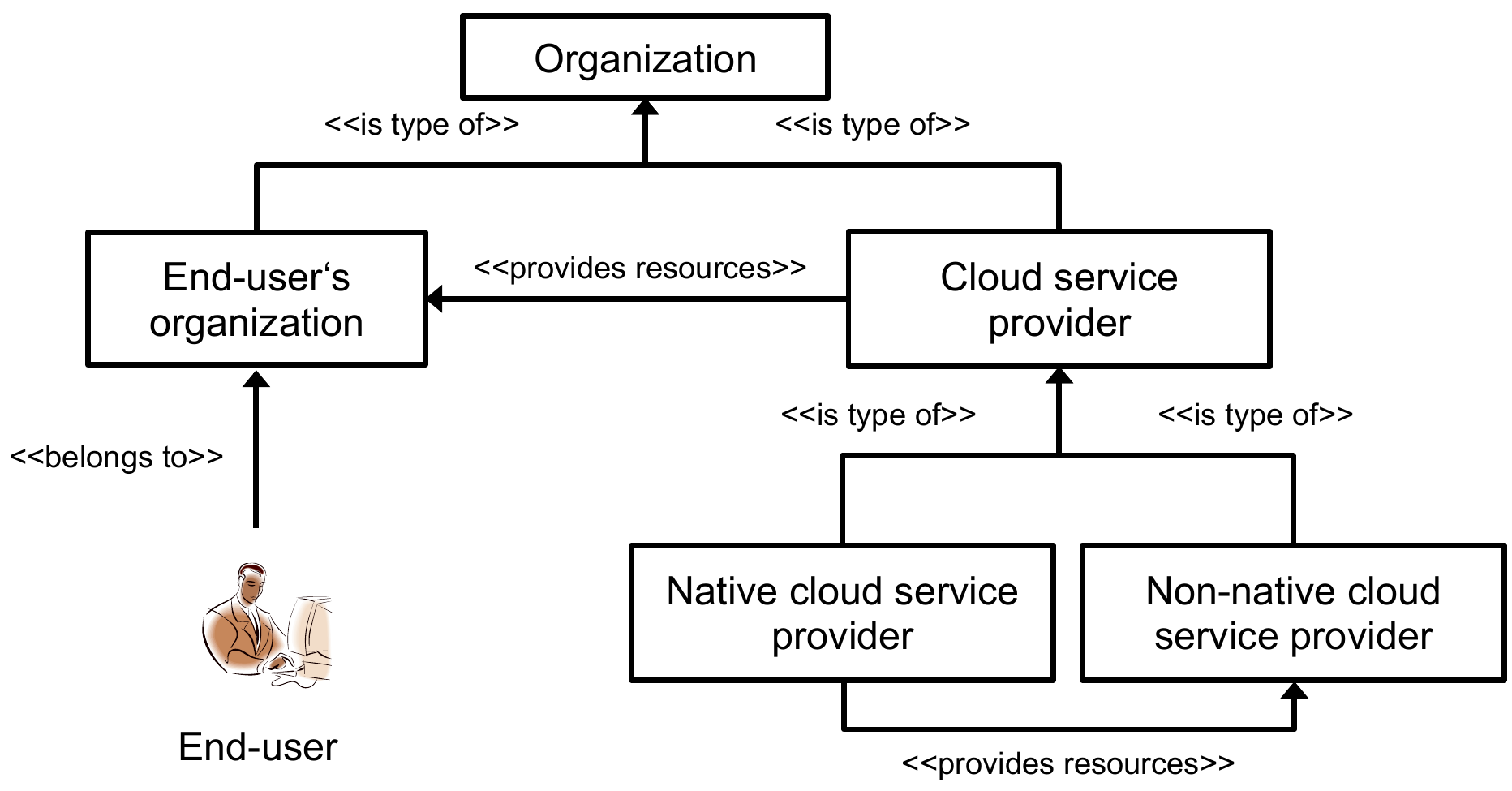} 
\end{center}
\caption{relationships between different stakeholders}
\label{fig:Organizations}
\end{figure}

We distinguish between the dimensions stakeholders, abstraction levels, roles, service-level agreements (SLAs), and size/volume: 

\vspace{5mm}

\emph{\textbf{Stakeholders}}: A stakeholder is defined as an entity that plays a certain \emph{role} in a given cloud usage scenario. We distinguish between two types of stakeholders: \emph{end-user} and \emph{organization}. An end-user is an individual or a programmed system that consumes cloud services (i.e., it uses provisioned resources). An organization is a legal entity that may either be the organization to which an end-user belongs or a \emph{cloud service provider} (i.e., an organization that provides cloud services). We distinguish between \emph{native} and \emph{non-native} cloud service providers. A native cloud service provider is a provider that \emph{owns} a cloud infrastructure, e.g., a data center. In contrast, a non-native cloud service provider does not own a cloud infrastructure, but it rather relies on provisioned resources from one or more native cloud service provider(s) in order to provide services to end-users. For instance, as a non-native cloud SaaS provider, we consider a company that provides an application to end-users according to the SaaS service delivery model where the application is deployed on leased infrastructure and/or platform resources from a native cloud provider. We depict the relationships between the considered stakeholders in Figure~\ref{fig:Organizations}. 

\vspace{5mm}

\emph{\textbf{Abstraction Levels}}: This dimension captures the levels at which the resource provisioning mechanisms of cloud providers abstract and manage the provisioned infrastructure, platform, or software resources in a given cloud environment. We differentiate between the abstraction levels \emph{hardware resources}\footnote{Under hardware resources, we understand the bare hardware (e.g., computing, storage, and network resources) including virtualization technology (if used). Note that the management and provisioning of hardware resources may be performed \emph{with} or \emph{without} the use of \emph{virtualization} technology. Although the bare hardware does not have any mechanisms for provisioning resources (in the form of cloud services), for the sake of completeness and consistency, we consider the hardware resources layer as a separate abstraction level in our resource provisioning hierarchy.}, \emph{IaaS}, \emph{PaaS}, and \emph{SaaS}. We assume that the resource provisioning mechanisms may be accessed through specialized access-enabling mechanisms (e.g., APIs), and may function in an independent manner. We also assume that the resource provisioning mechanisms enable resource provisioning between providers at different abstraction levels. In that case, the involved providers at different abstraction levels interact in a strict hierarchical relationship; that is, providers at a lower abstraction level may provide services to a provider at a higher abstraction level, but not vice-versa. We consider SaaS as the highest abstraction level, hardware resources as the lowest, with the PaaS and IaaS abstraction levels in between. Given the previously mentioned hierarchy of abstraction levels, an example resource provisioning relationship between providers at different abstraction levels is an IaaS provider providing infrastructure resources to a PaaS provider, but not vice-versa\footnote{An exception of the strict hierarchical order in the resource provisioning hierarchy exists in the case where a provider at a given abstraction level provisions resources to a provider at the same abstraction level (e.g., in cases where a mediator is involved in the resource provisioning to add value), a topic discussed later in Section~\ref{sec:patterns:textual}.}.

\vspace{5mm}

\emph{\textbf{Roles}}: As we mentioned previously when discussing stakeholders, we assume that any stakeholder in a cloud usage scenario plays a given \emph{role}. We differentiate between the roles \emph{provider}, \emph{consumer}, and \emph{intermediary}. A provider provides resource provisioning services to consumers and may be a native or a non-native cloud provider that operates at any of the abstraction levels Hardware resources, IaaS, PaaS, or SaaS. We do not consider an end-user as a provider in any situation. A consumer may either be a provider operating at one of the abstraction levels IaaS, PaaS, or SaaS, and consuming resources at the same time, or an end-user. We refer to a provider at any abstraction level that consumes and provides resources at the same time as an \emph{intermediary}. For instance, a SaaS provider provisioning a software application to end-users and using platform resources leased from a PaaS provider at the same time, is an intermediary.

\begin{figure}[ht]
\begin{center}
\includegraphics[scale=0.7]{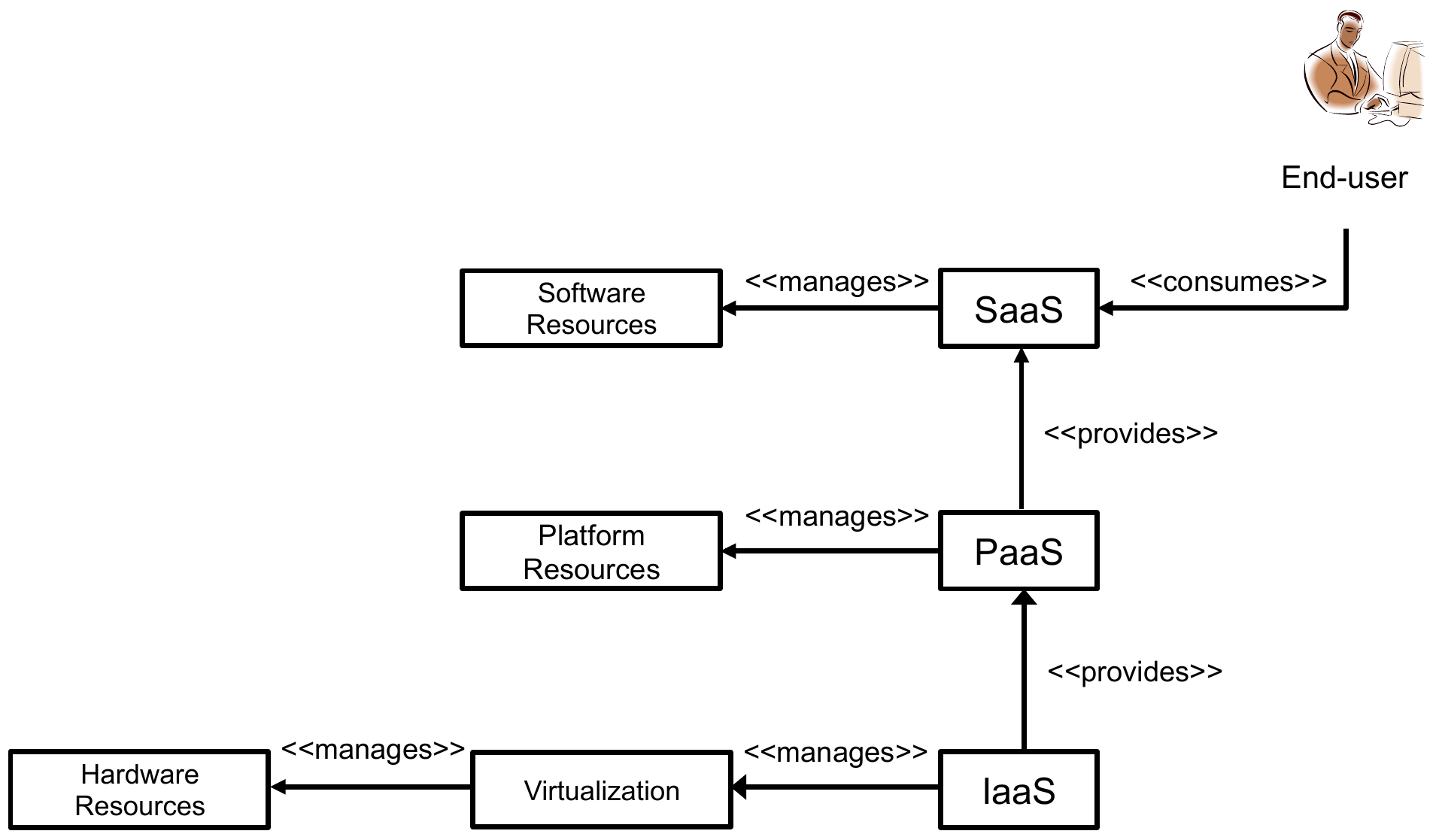} 
\end{center}
\caption{interactions between stakeholders in a resource provisioning scenario 
}
\label{fig:levelsroles}
\end{figure}

According to the previously presented resource provisioning hierarchy between abstraction levels, providers at lower abstraction levels may provision resources to consumers at higher abstraction levels, but not vice versa. In Figure~\ref{fig:levelsroles}, we depict the interactions between stakeholders in the case where an end-user consumes software resources from a SaaS provider that uses virtualized infrastructure and platform resources provisioned from IaaS and PaaS providers, respectively, according to the hierarchy which applies in such a provisioning.  

\vspace{5mm}

\emph{\textbf{Service Level Agreements (SLAs)}}: We distinguish between \emph{internal} and \emph{external} SLAs. \emph{Internal} SLAs define QoS requirements and customer contracts for resource provisioning scenarios within the jurisdiction of a \emph{single} organization. For instance, the QoS requirements related to the provisioning of infrastructure resources from an IaaS provider to an intermediary at the PaaS abstraction level, in the case where both are within the legal jurisdiction of a single cloud provider, are defined as part of an internal SLA. In contrast, \emph{external} SLAs define QoS requirements and customer contracts for resource provisioning scenarios where different organizations are involved. For instance, the QoS requirements for the provisioning of software resources from a SaaS provider to an end-user, who does not belong to the SaaS provider's organization, are defined as part of an external SLA. The notions of internal and external SLAs can be used to differentiate between private and public cloud infrastructures. A cloud usage scenario where the QoS requirements and customer contracts related to the resource provisioning processes between all involved (provider, consumer) pairs are defined as part of internal SLAs is a usage scenario of a private cloud environment.

\vspace{5mm}

\emph{\textbf{Size/Volume}}: The size/volume dimension is a quantifier of the amount of provisioned resources from a provider to a consumer in a given resource provisioning scenario. In the following Section~\ref{sec:elementarycups}, we present the formal notation of this dimension in greater detail.

\subsection{Textual and Visual Cloud Usage Patterns}
\label{sec:textual_and_visual_formalism}

Taking into account the relevant dimensions described in Section~\ref{sec:dimensions:cups}, we now present a formalism for formal description of real-world cloud usage scenarios. The proposed formalism has been designed to satisfy the following requirements (RQs):

\vspace{5mm}

\textbf{RQ 1.} \emph{Expressiveness}: A cloud usage pattern should convey information covering all relevant dimensions of the described cloud usage scenario. The formalism should be expressive enough to describe any common cloud usage scenario that occurs in practice.

\vspace{5mm}

\textbf{RQ 2.} \emph{Mutual Exclusiveness}: A given cloud usage scenario should fit into the description of at most one cloud usage pattern.

\vspace{5mm}

\textbf{RQ 3.} \emph{Comprehensibility}: A cloud usage pattern should be intuitive and easy to understand.

\vspace{5mm}

\textbf{RQ 4.} \emph{Determinism}: The process for defining a cloud usage pattern should be clearly defined and produce deterministic results. 

\vspace{5mm}

Given the above requirements, we introduce two forms of cloud usage patterns: textual and visual.

\vspace{5mm}

\subsubsection{Textual Form of Cloud Usage Patterns}
\label{sec:textual-cloud-usage-patterns}
\label{sec:patterns:textual}

We distinguish between: 

\begin{quote}
(i) \emph{elementary} cloud usage patterns for describing cloud usage scenarios where a single provider at any abstraction level provisions resources to a consumer, i.e., an intermediary or an end-user. An example would be the typical scenario where a single IaaS provider provisions infrastructure resources to a SaaS provider for hosting a software application provisioned to end-users.
\end{quote}

\begin{quote}
(ii) \emph{extended} patterns for describing describing more complex cloud usage scenarios, e.g., hybrid resource provisioning (i.e., provisioning of resources from multiple providers at the same abstraction level to a consumer) and value chains with mediators. For instance, an example of the former scenario is the case where two IaaS providers provision infrastructure resources to a SaaS consumer for hosting a software application provisioned to end-users. An example of the latter scenario is the case of an organization specializing in the operation and management of the resources provided by given PaaS provider, consumes platform resources and re-provisions them to end-users. 
\end{quote}

In Section~\ref{sec:practice:services}, we present several real-world scenarios similar to the above mentioned examples. Next, we discuss the textual form of elementary cloud usage patterns.  

\paragraph{Elementary Cloud Usage Patterns}
\label{sec:elementarycups}

\begin{figure}[h] 
\begin{center}
\includegraphics[scale=0.8]{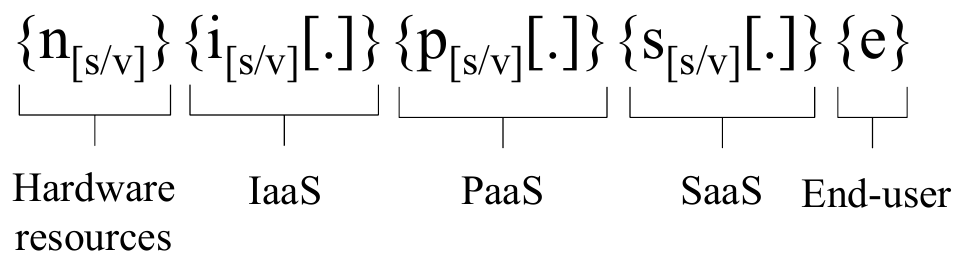} 
\end{center}
\caption{structure of an elementary cloud usage pattern in textual form}
\label{fig:patternssections}
\end{figure}

An elementary cloud usage pattern, in its textual form, is specified as a string consisting of several sections where each section corresponds to an abstraction level: \emph{Hardware resources}, \emph{IaaS}, \emph{PaaS}, \emph{SaaS}, or \emph{End-user}. Next, we present the rules that define the syntax of the proposed notation:

\vspace{3mm}

\textbf{Rule E.I)}  In Figure~\ref{fig:patternssections}, we depict the structure of an elementary cloud usage pattern in textual form.

\begin{quote}
\textbf{Rule E.I.1)} When read form left to right, the sections must follow the exact order shown in Figure~\ref{fig:patternssections}, however, only those sections must be included that apply to the considered scenario.
\end{quote}

\begin{quote}
\textbf{Rule E.I.2)} A pattern must consist of at least two sections, including the section \emph{End-user}, and at least one of the sections \emph{IaaS}, \emph{PaaS}, or \emph{SaaS}. The section \emph{Hardware resources} may only be used in cases where  hardware resources are provisioned without the use of virtualization technology.
\end{quote} 

\begin{quote}
\textbf{Rule E.I.3)}  The section \emph{End-user} may contain at most one character. The sections \emph{Hardware resources}, \emph{IaaS}, \emph{PaaS}, and \emph{SaaS} may contain at least one character, and at most two characters and one number.
\end{quote}

\begin{quote}
\textbf{Rule E.I.4)} The provisioning size/volume at any abstraction level (depicted as \texttt{[s/v]} in Figure~\ref{fig:patternssections}) is a quantifier specified using numerical notation. The specification of the provisioning size/volume at any abstraction level is optional. For instance, such a specification may be ommited when one does not have the respective information, or when the provisioning size/volume is of no importance in the considered scenario. 
\end{quote}

\begin{quote}
\textbf{Rule E.I.5)} At maximum one number may be used to specify the provisioning size/volume at any abstraction level. The number that specifies the provisioning size/volume at a given abstraction level is a subscript placed next to the letter that denotes the respective abstraction level (i.e., \texttt{n}, \texttt{i}, \texttt{p}, or \texttt{s}).
\end{quote}

\begin{quote}
\textbf{Rule E.I.6)} The character ``dot'' (\texttt{.}) may be included only  in the sections \emph{IaaS}, \emph{PaaS}, or \emph{SaaS}, however, only in the sections that apply to the considered scenario.
\end{quote}

Note that the capitalization of letters in a cloud usage pattern does not convey any specific information; that is, one may use uppercase and/or lowercase letters. However, we recommend the use of consistent letter capitalization for clarity and consistency. Next, we present the rules that define the semantics of the proposed notation:

\vspace{4mm}

\textbf{Rule E.II)} The semantics of the characters that may be used when specifying an elementary cloud usage pattern in textual form are defined in Table~\ref{tab:meaning}. 

\begin{table}[h]
\caption{semantics of the characters used in cloud usage patterns.}
\centering
\begin{tabular}{|p{2cm}|p{12cm}|}
\hline
 Character & Meaning \\
\hline 
\hline
\texttt{n}  & Denotes hardware provisioning without the use of \emph{virtualization} technology.\\
\hline
\texttt{i}  & Denotes a cloud provider/intermediary at the \emph{IaaS} abstraction level.\\
\hline
\texttt{p} & Denotes a cloud provider/intermediary at the \emph{PaaS} abstraction level.\\
\hline
\texttt{s} & Denotes a cloud provider/intermediary at the \emph{SaaS} abstraction level.\\
\hline
\texttt{.} &  Denotes an \emph{external} SLA as part of which the QoS requirements and customer contracts for a resource provisioning relationship between a provider and a consumer belonging to two different organizations are defined, i.e., it denotes crossing of a legal boundary between two different organizations in a resource provisioning relationship.\\
\hline
\texttt{e} & Denotes an \emph{end-user}.\\
\hline
\end{tabular}
\label{tab:meaning}
\end{table}

\begin{quote}
\textbf{Rule E.II.1)} The order of the letters, when read from left to right, denotes the (\emph{provider, consumer}) pairs that exist in a given cloud usage scenario. For instance, the pattern \texttt{ips.e} denotes the following (provider, consumer) pairs: (\emph{Hardware resources, IaaS}), (\emph{IaaS, PaaS}), (\emph{PaaS, SaaS}), and (\emph{SaaS, End-user}). Consequently, a letter that has an adjacent letter or the character ``dot''(.) on both sides, denotes an \emph{intermediary}. In the above pattern, the providers at the abstraction levels \emph{PaaS} and \emph{SaaS} denoted by the letters \texttt{p} and \texttt{s} are intermediaries.
\end{quote}

\begin{quote}
\textbf{Rule E.II.2)} Two adjacent letters that \emph{are} separated by a ``dot'' (.) indicate that the QoS requirements and customer contracts for the resource provisioning relationship in the respective (provider, consumer) pair are defined as part of an \emph{external} SLA. Consequently, two adjacent letters that \emph{are not} separated by a ``dot'' (.) imply the opposite situation. For instance, the pattern \texttt{ips.e} indicates that the QoS requirements and customer contracts for the resource provisioning relationships (\emph{IaaS, PaaS}) and (\emph{PaaS, SaaS}) are defined as part of an internal SLA, whereas the requirements and contracts for the resource provisioning relationship (\emph{SaaS, End-user}) are defined as part of an external SLA. 
\end{quote}

\begin{quote}
\textbf{Rule E.II.3)} It is assumed that the hardware resources are provisioned \emph{with} the use of virtualization technology, which is the most common case for modern cloud systems. For this reason, the use of virtualization is not specified explicitly in the proposed notation, however, in the unlikely case that the hardware is provisioned \emph{without} the use of virtualization, this can be specified by including the character \texttt{n} in section \emph{Hardware resources}, which is normally omitted. For instance, the pattern \texttt{ns.e} specifies that the hardware used by the \emph{SaaS} abstraction level is provisioned without the use of virtualization, whereas the contrary applies for the pattern \texttt{s.e}.
\end{quote}

\begin{quote}
\textbf{Rule E.II.4)} The resource provisioning size/volume specified in the dimension size/volume (denoted as \emph{[s/v]} in Figure~\ref{fig:patternssections}) in any of the sections \emph{Hardware resources}, \emph{IaaS}, \emph{PaaS}, or \emph{SaaS}, is a specification of a metric indicating the size/volume of provisioned resources at the respective levels. The provisioning size/volume may be specified in an arbitrary unit of measurement which has to explicitly defined, for example, an order of magnitute (e.g., hundreds, thousands), or a logarithmic value.  
\end{quote}

\vspace{5mm}

The specification rules for elementary cloud usage patterns given above provide a formal notation for almost all of the relevant dimensions of information and their categories described in Section~\ref{sec:dimensions:cups}. The only dimension that is not covered by the above notation is the cloud service provider category. This category consists of the sub-categories native and non-native cloud service providers. As defined in Section~\ref{sec:dimensions:cups}), under native cloud service provider, we understand a provider that owns a cloud infrastructure and provisions infrastructure, platform, and/or software resources. In a cloud usage pattern the provider that operates at the abstraction level of the lowest order is a native cloud service provider and it provisions resources at the abstraction levels that are specified up until the first ``dot'' (.) when reading the pattern from left to right. The resource provisioning at the other abstraction levels is performed by non-native cloud providers. For instance, the pattern \texttt{ip.s.e} describes a scenario in which a native cloud provider provisions infrastructure and platform resources, and a non-native cloud provider provisions software resources to an end-user.

\paragraph{Cloud Usage Patterns for Hybrid Services}
\label{sec:cupdsforhybrid}

With the elementary cloud usage patterns defined, we extend the existing formalism to support the formal description of hybrid resource provisioning scenarios which are common in practice. Under hybrid resource provisioning, we understand the provisioning of resources to a consumer (i.e., a provider at a given abstraction level or an end-user) from multiple native or non-native cloud providers at the same abstraction level. For instance, when the capacity of a platform-level provider is reached, additional platform resources may be leased from another platform provider. In order to support the specification of hybrid services in our formalism, we introduce the following rules:

\begin{quote}
\textbf{Rule H.I)} The specification of multiple different providers provisioning resources at the same abstraction level is performed using similar notation as the notation for elementary cloud usage patterns (Rule H.II) where for each provider the respective spectification is enclosed in parentheses. An example would be the pattern \texttt{(ip)(i.p.)s.e}, in which the two specifications in parentheses \texttt{ip} and \texttt{i.p.} correspond to two independent providers of platform resources. The specifications in parentheses may contain nested specifications of the same nature, also enclosed in parentheses. Further, the order in which specifications at the same nesting level are written is irrelevant, for example, the patterns \texttt{(ip)(i.p.)s.e} and \texttt{(i.p.)(ip)s.e} are equivalent.
\end{quote}

\begin{quote}
\textbf{Rule H.II)} The specifications enclosed in parentheses must conform to the rules for elementary cloud usage patterns with the exception that they are not required to contain all mandatory sections (e.g., the section \emph{end-user}). To the contrary, such specifications normally only contain sections up to the abstraction levels at which the providers involved in the respective hybrid resource provisioning scenario operate (i.e., beginning with the lowest applicable abstraction level and ending at the abstraction level at which multiple providers provide resources of the \emph{same} type to a consumer). For instance, in the pattern \texttt{(ip)(i.p.)s.e}, both specifications \texttt{(ip)} and \texttt{(i.p.)} end by specifying the PaaS abstraction level with the letter \texttt{p}, indicating that multiple platform providers jointly provision platform resources to a consumer at the SaaS abstraction level.
\end{quote}

\begin{quote}
\textbf{Rule H.III)} In the case where the QoS requirements and customer contracts for a given (provider, consumer) pair are defined as part of an external SLA, and where the provider is involved in a hybrid resource provisioning scenario, the external SLA is specified by placing a``dot'' (.) character \emph{inside} the parentheses that enclose the specification corresponding to the respective provider. For instance, the specification \texttt{(i.p.)} in the pattern \texttt{(ip)(i.p.)s.e}, denotes a platform provider that is not within the same legal boundaries as the consumer at the SaaS abstraction level. To the contrary, the specification \texttt{(ip)} indicates that a second platform provider is used that is within the same legal boundaries as the consumer at the SaaS abstraction level. 
\end{quote}

\begin{quote}
\textbf{Rule H.IV)} The total size/volume of the provided/consumed resources in a hybrid resource provisioning scenario is equal to the sum of the resource provisioning sizes/volumes of each involved provider. For instance, given the pattern \texttt{(ip$_1$)(i.p$_2$.)s.e} where the provisioning size/volume is specified in the unit of hundreds, one may assume that the consumer at the SaaS abstraction level consumes 300 resource units in total. 
\end{quote}

\paragraph{Cloud Usage Patterns for Value Chains with Mediators}
\label{sec:cupsforvaluechains}

As discussed in Section~\ref{sec:cloud-usage-patterns}), the term value chain, in the context of cloud computing, is normally used to refer to a network of multiple service providers that cooperate in order to add/generate value to a consumer \cite{leimeister+2010}. The presented formalism for elementary and hybrid cloud usage patterns allows to specify different interrelation scenarios involving multiple cloud providers in a value chain (i.e., scenarios where a single provider provisions resources at a single abstraction level in a hierarchical order of multiple such levels, and where multiple providers provision resources at a single abstraction level, also in a hierarchical order of multiple such levels). In this section, we extend our formalism to support the specific case where an organization that does not own a cloud infrastructure including infrastructure, platform, or software resources, leases such resources provisioned from a single, or multiple, native or non-native cloud IaaS, PaaS, or SaaS providers, in order to re-provision them to a consumer with added value. In the context of this work, we refer to such an organization as a \emph{mediator}.  In the commercial cloud market, mediators are also commonly referred to as value-adding resellers (VARs). The trend having mediators involved in cloud resource provisioning has only recently emerged and it is attracting a significant amount of attention. For instance, many native cloud providers have reseller programs enabling cost- and time-efficient provisioning of infrastructure, platform, and/or software resources to mediators for reselling. An example is the reseller program of the Google App software provider, which has attracted over 6000 authorized resellers \cite{googleapp:resellers}. 

The value added by mediators to resource provisioning services may be of various nature, for example, a mediator might possess strong expertise in a relevant domain (e.g., provisioning mechanisms and policies such as resource allocation, task scheduling, and similar) adding significant value to the resource provisioning services through enhanced efficiency. Further, a mediator may provision resources at a lower cost than the native and/or non-native cloud provider(s) from which it leases resources; that is, in the case where a mediator is a high-volume customer of its cloud provider(s), it may be eligible for discounts allowing to resell resources at lower prices\footnote{An example of a commercial native cloud provider that offers discounts for high-volume customers is Amazon \cite{amazon:resourcediscount}.}. 

The specific class of value chains considered in this section (i.e., value chains with mediators) may involve resource provisioning from a single native or non-native cloud provider, or from multiple such providers, for reselling by a mediator. Given the latter scenario, we extend the previously defined rules for specifying hybrid services (Section~\ref{sec:cupdsforhybrid}) to support value chains with mediators. 

\begin{quote}
\textbf{Rule M.I)} A mediator relying on leased resources from a \emph{single} cloud provider at a given abstraction level is denoted using the letter that corresponds to the specific abstraction level at which resources are provisioned for reselling, i.e., \texttt{n} (optional, otherwise no letter is used, see Rule E.I.2), \texttt{i}, \texttt{p}, or \texttt{s}. This letter is written after the pattern specification for the provider from which resources are leased enclosed in parentheses, and which is similar to an elementary cloud usage pattern (see Rule H.I and Rule H.II). An example is the pattern \texttt{(i.)i.e}, where the second \texttt{i}, when one reads from left to right, denotes a mediator provisioning infrastructure resources to an end-user, leased from a single infrastructure-level cloud provider.
\end{quote}

\begin{quote}
\textbf{Rule M.II)} A mediator relying on leased resources from \emph{multiple} independent providers at the same abstraction level is denoted using the letter that corresponds to the specific abstraction level at which resources are provisioned for reselling, i.e., \texttt{n} (optional, otherwise no letter is used, see Rule E.I.2), \texttt{i}, \texttt{p}, or \texttt{s}. This letter is written after the pattern specifications for the providers from which resources are leased enclosed in parentheses, i.e., \texttt{()}, and which are similar to elementary cloud usage patterns (see Rule H.I and Rule H.II). An example is the pattern \texttt{(ip.)(i.p.)p.s.e}, where the third \texttt{p}, when one reads from left to right, denotes a mediator that provisions leased platform resources from two platform providers, to a consumer at the SaaS abstraction level.  
\end{quote}

\begin{quote}
\textbf{Rule M.III)} As for the elementary cloud usage patterns, the character ``dot'' (.) denotes an external SLA (i.e., in the context of value chains it indicates that the provider provisioning resources for reselling and the mediator are separate legal entities). Note that the use of a ``dot'' (.) is mandatory, since by definition a mediator is an independent legal entity that leases and resells resources from a single or multiple native or non-native cloud providers.    
\end{quote}

\begin{quote}
\textbf{Rule M.IV)} The specification of the provisioning size/volume of a mediator is optional (see similar rule E.I.4). 
\end{quote}

\subsection{Visual Form of Cloud Usage Patterns}
\label{sec:graphical-cloud-usage-patterns}
\label{sec:patterns:visual}

In this section, we propose a visual formalism depicting cloud usage patterns in a graphical notation. The visual formalism is not intended to be as strict as the textual formalism we introduced in the previous section; that is, we consider it merely as a recommendation on visualizing textual forms of cloud usage patterns in an intuitive manner and in a form easy to understand by a wide audience. Also, we do not claim improved features and or higher expressive power over existing visual formalisms for value chains, such as the e$^3$-method proposed by Gordjin \cite{Gordijn02}. To the contrary, our formalism is intended for straightforward use and immediate interpretation by a wide audience, whose members are not necessarily assumed to have extensive expert knowledge on value chains, especially in the context of cloud computing. 

The proposed visual formalism is based on the graphical elements presented in Table~\ref{tab:graphical-elements} and Table~\ref{tab:graphical-elements-extended}. The graphical elements listed in Table~\ref{tab:graphical-elements} may be used to visualize elementary cloud usage patterns (Section~\ref{sec:dimensions:cups}), while the graphical elements listed in Table~\ref{tab:graphical-elements-extended} allow to visualize extended cloud usage patterns describing hybrid services and value chains with mediators. 

\begin{table}[ht]
\centering
\caption{graphical elements for specifying elementary cloud usage patterns in visual form
}
\begin{tabular} { |m{4cm} | m{6.5 cm} | m{3cm}| } \hline
\multicolumn{2}{|l|}{Dimension} & \multicolumn{1}{|l|}{Graphical Element}\\ \hline
\hline

\vspace{0.05pt} Abstraction Levels &  \vspace{0.005pt}SaaS & \vspace{0.005pt} \includegraphics[scale=0.5]{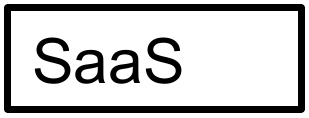}  \\ 
& PaaS & \includegraphics[scale=0.5]{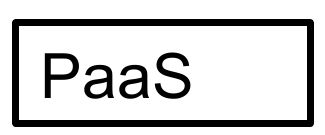}   \\ 
& IaaS & \includegraphics[scale=0.5]{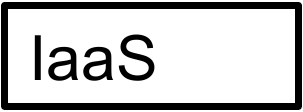}   \\ 
& Virtualization\footnotemark & \includegraphics[scale=0.5]{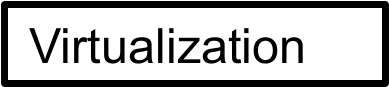} \\ 
& Hardware resources& \includegraphics[scale=0.5]{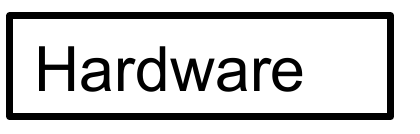} \\ 
\hline

\vspace{0.005pt} Stakeholders & \vspace{0.005pt} Native cloud service provider & \vspace{0.005pt} \includegraphics[scale=0.6]{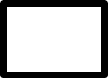} \\ 
& Non-native cloud service provider & \includegraphics[scale=0.6]{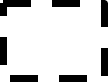}  \\
& End-user's organization & \includegraphics[scale=0.6]{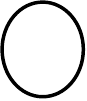}   \\ 
 & End-user & \includegraphics[scale=1]{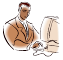}  \\ 
\hline

\vspace{0.005pt}SLAs & \vspace{0.005pt}Internal & \vspace{0.005pt}\includegraphics[scale=0.55]{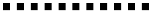}  \\
& External & \includegraphics[scale=0.55]{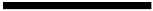}   \\\hline

\vspace{0.005pt}Roles & \vspace{0.005pt}Consumer & \vspace{0.005pt}\includegraphics[scale=0.15]{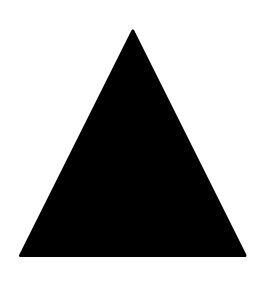} \\
& Provider & \includegraphics[scale=0.15]{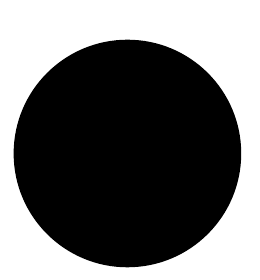}   \\\hline

\vspace{0.005pt}Size/Volume (optional) & \vspace{0.005pt}Provisioning with internal SLA & \vspace{0.005pt}\includegraphics[scale=0.85]{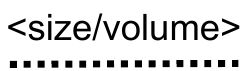} \\
&\vspace{0.005pt}Provisioning with external SLA & \vspace{0.005pt}\includegraphics[scale=0.85]{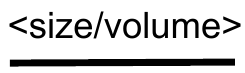} \\\hline

\end{tabular}
\label{tab:graphical-elements}
\end{table}

\begin{table}[ht]
\centering
\caption{graphical elements for specifying extended cloud usage patterns in visual form
}
\begin{tabular} { |m{4cm} | m{9cm}| } \hline
Graphical Element & Meaning\\ \hline
\hline

\includegraphics[scale=0.5]{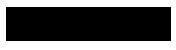}  & Hybrid resource provisioning (merged resources)\\
\includegraphics[scale=0.5]{Figures/GraphicalElements/TraditionalServiceProvider.pdf}  & Mediator (non-native cloud service provider)
\\\hline

\end{tabular}
\label{tab:graphical-elements-extended}
\end{table}

\begin{figure}[ht]
\begin{center}
\includegraphics[scale=0.58]{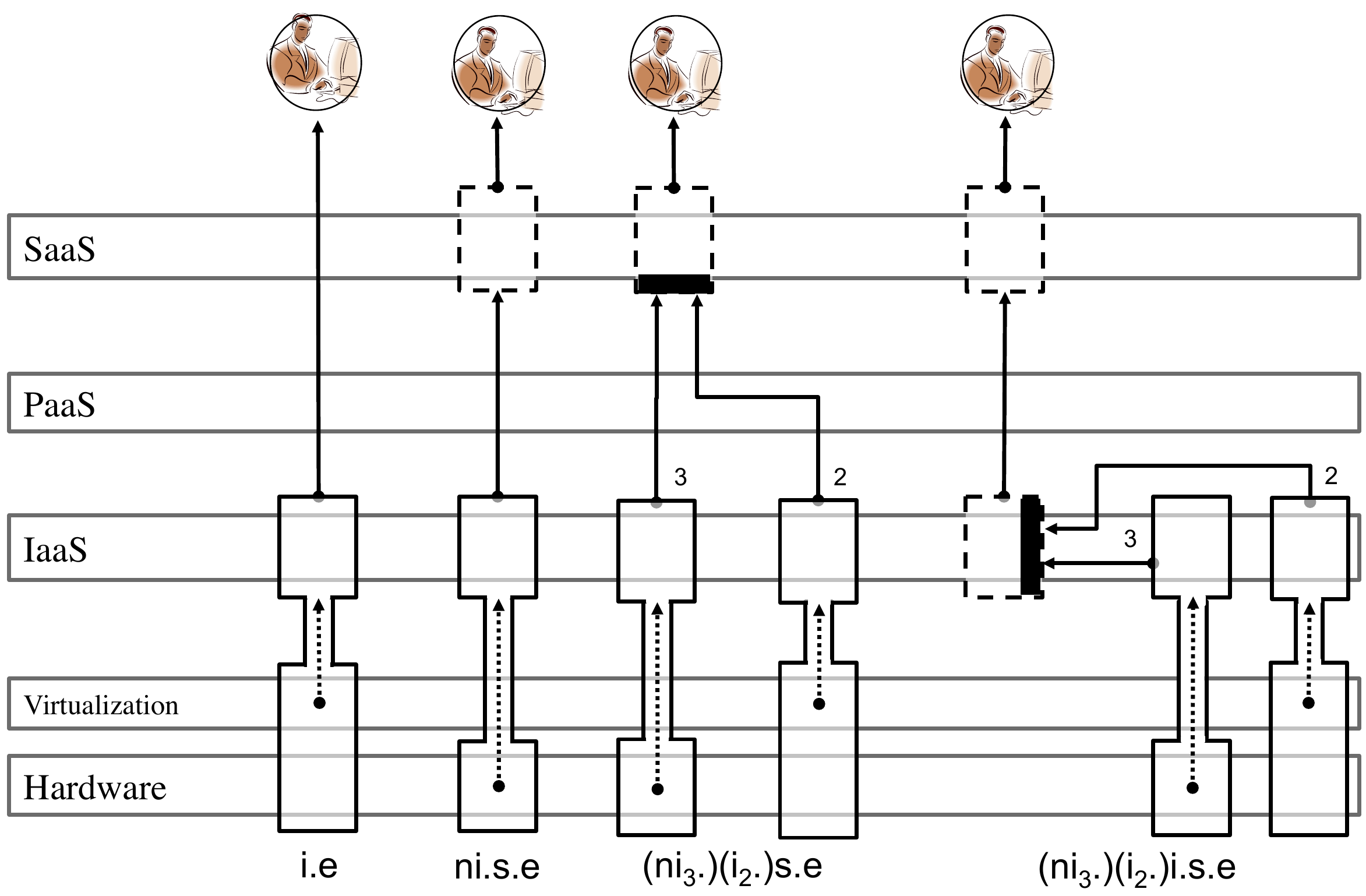} 
\end{center}
\caption{visual form of the textual cloud usage patterns \texttt{i.e}, \texttt{ni.s.e}, \texttt{(ni$_{3}$.)(i$_{2}$.)s.e}, and \texttt{(ni$_{3}$.)(i$_{2}$.)i.s.e}. 
}
\label{fig:CloudPatternsExample1}
\end{figure}

In Figure~\ref{fig:CloudPatternsExample1}, we present several examples of the use of the graphical elements listed in Table~\ref{tab:graphical-elements} and Table~\ref{tab:graphical-elements-extended} for visualizing elementary (i.e., \texttt{ni.s.e}) and extended cloud usage patterns (i.e., \texttt{(ni$_{3}$.)(i$_{2}$.)i.s.e}). 

The pattern \texttt{i.e} describes a common situation where a native IaaS cloud provider provisions infrastructure resources to an end-user. The native cloud provider, depicted as a box with solid line, and the end-user belong to different organizations. Thus, the QoS requirements and customer contracts are defined as part of an external SLA visualized with a solid line. The underlying hardware resources and the IaaS abstraction level belong to the same organization and therefore the QoS requirements and customer contracts for the provisioning of hardware resources are defined as part of an internal SLA visualized with a dashed line. The hardware resources are provisioned with the use of virtualization technology. This is reflected by placing the graphical symbol of a provider inside the box representing the virtualization abstraction level. An important difference between the pattern \texttt{ni.s.e}, also depicted in Figure~\ref{fig:CloudPatternsExample1}, and the pattern \texttt{i.e} is that in the former hardware resources are provisioned without the use of virtualization technology. Therefore, the graphical symbol of a provider is placed inside the box representing the hardware resources abstraction level. Further, the pattern \texttt{ni.s.e} describes provisioning of software resources from a non-native cloud SaaS provider, which is therefore visualized using a box with dashed line.

\footnotetext{In the context of this work, virtualization is considered as part of the abstraction level Hardware resources (Figure~\ref{fig:DimensionsCategories}). However, in Table~\ref{tab:graphical-elements}, we list virtualization as a separate abstraction level for clarity of the presentation. For the same reason, we refer to it as ``the virtualization abstraction level''.}

Beside the elementary patterns \texttt{i.e} and \texttt{ni.s.e}, we also visualize the pattern \texttt{(ni$_3$.)(i$_2$.)s.e}, describing hybrid provisioning of infrastructure resources to a consumer at the SaaS abstraction level, and the pattern \texttt{(ni$_3$.)(i$_2$.)i.s.e}, describing the involvement of a mediator in the hybrid resource provisioning. The considered hybrid resource provisioning scenario is a scenario where a consumer at the SaaS abstraction level consumes infrastructure resources from two infrastructure providers conforming to the patterns \texttt{ni.s.e} and \texttt{i.e}, respectively. Thus, to visualize this scenario, we apply the visual forms of the elementary patterns \texttt{i.e} and \texttt{ni.s.e} as a basis and use a thick line as a graphical element to visualize the hybrid provisioning of infrastructure resources from the two infrastructure providers. We place this line inside the graphical element for a non-native cloud service provider at the abstraction level at which the provisioned resources from the two infrastructure providers are merged (i.e., SaaS). A thick line is used in many other visual formalisms for similar purposes, i.e., for depiction of common activities of multiple stakeholders, a prominent example being the activity diagrams in the UML (Unified Modeling Language) \cite{umlspecification}. Note that an alternative visualization of a hybrid resource provisioning is also possible in the case where one of the multiple providers involved in a hybrid resource provisioning abstracts the resources provisioned by the other providers and provisions them to a consumer. This is usually the case when the provider mentioned above is within the same legal domain as the consumer. An example of such a case is the scenario described by the textual pattern \texttt{(i.)(i)s.e}, where a private IaaS provider expands (i.e., ``spills over'') to a public IaaS provider when its capacity is reached in order to provision a sufficient amount of resources to the consumer at the SaaS abstraction level. From the consumer's perspective, the provisioning of infrastructure resources is performed by a single IaaS provider, that is, the private IaaS provider. When visualizing the above scenario, a thick line is placed inside the graphical element visualizing the private infrastructure provider to indicate the merging of the private and public infrastructure resources at the IaaS abstraction level. We present a real-world example of such a scenario and visualization of the respective pattern in Section~\ref{sec:practice:patterns}.

In the visual forms of the patterns \texttt{(ni$_3$.)(i$_2$.)s.e} and \texttt{(ni$_3$.)(i$_2$.)i.s.e}, we also specify the provisioning sizes/volumes by placing the respective numbers next to the lines that visualize the SLAs and close to the graphical elements that visualize the providers whose provisioning size/volume is specified.  

In order to visualize a value chain with a mediator, we use the same set of graphical elements used when visualizing hybrid resource provisioning without a mediator, however, we use in addition to this the graphical element for a non-native cloud service provider (i.e., a mediator), which in this case consumes resources from two infrastructure providers and operates at the same abstraction level as the infrastructure providers (i.e., IaaS). Note that in the depicted example, the mediator uses hybrid resource provisioning to provision infrastructure resources to a consumer at the SaaS abstraction level, thus the use of the thick line. We provide an example in which a mediator does not use hybrid resource provisioning in Section~\ref{sec:practice:patterns}, where we provide examples of real-world cloud usage scenarios.

\newpage 

\section{Cloud Usage Patterns in Practice}\label{sec:practice}

In this section, we present examples of how cloud usage patterns can be used to describe real-world cloud usage scenarios. To this end, in Section~\ref{sec:practice:services}, we provide a brief overview of several cloud usage scenarios, and then, in Section~\ref{sec:practice:patterns}, we apply our formalism to formally describe these scenarios. 

\subsection{Real-World Cloud Usage Scenarios}\label{sec:practice:services}




In order to identify practically relevant cloud usage scenarios, we surveyed multiple reports on resource provisioning from commercial cloud providers that play a major role in the cloud computing market, such as Facebook, Amazon, and similar. 
 
\vspace{5mm}

\textit{\textbf{Scenario AWS}}: \textit{Amazon Web Services} \cite{AmazonWebServices} is a native cloud provider that provisions infrastructure resources to end-users. This includes provisioning of computing, database, storage, and networking resources offered in the form of ``Web Services''. In Table~\ref{tab:Amazon}, we list some of these services as well as the respective types of provisioned infrastructure resources \cite{aws:documentation}. 

\begin{table}[ht]
\caption{several Amazon Web Services}
\centering
\begin{tabular}{|p{7.5cm}|p{6.5cm}|}
\hline
Amazon Web Service & Infrastructure Resources \\
\hline 
\hline
Amazon Elastic Compute Cloud (EC2)

Amazon Elastic MapReduce (EMR)  & Computing\\
\hline
Amazon Simple Storage Service (S3)

Amazon Elastic Block Store (EBS) & Storage\\
\hline
Amazon DynamoDB
    
Amazon Relational Database Service (RDS) & Database\\
\hline
Amazon Virtual Private Cloud (VPC)

Amazon Route53 & Networking\\
\hline
\end{tabular}
\label{tab:Amazon}
\end{table}


\vspace{5mm}

\textit{\textbf{Scenario FBK}}: At the time of writing, \emph{Facebook} \cite{Facebook} is the world's largest online social networking application. The Facebook application is delivered to end-users according to the SaaS service delivery model. It is deployed on a private cloud infrastructure, which does not use virtualization technology for maximum performance and efficiency \cite{ZDNetFacebook}.
The Facebook social networking application relies on platform resources provisioned by a PaaS provider residing on the same cloud infrastructure as the application. This provider also provisions resources to end-users; that is, it provides an API for developing applications supported by the cloud infrastructure in which it resides, and therefore, such applications may be integrated in Facebook itself. 



\vspace{5mm}

\textit{\textbf{Scenario GAN}}: \textit{Go!Animate}~\cite{GoAnimate} provides a web application for developing animation videos. For maximum efficiency and scalability, Go!Animate leases infrastructure resources from the native cloud infrastructure provider \textit{Amazon Web Services} for application hosting and management. The Amazon Web Services used by Go!Animate are listed in Table~\ref{tab:GoAnimateAmazon} \cite{GoAnimateAmazon}. 

\begin{table}[ht]
\caption{Amazon Web Services used by Go!Animate}
\centering
\begin{tabular}{|p{8cm}|p{6cm}|}
\hline
Amazon Web Service & Usage \\
\hline 
\hline
Amazon Elastic Compute Cloud  & Hosting of web servers\\
\hline
Amazon Relational Database Service  & Hosting of a relational database\\
\hline
Amazon Simple Storage Service  & Hosting of static content\\
\hline
Amazon Simple Queue Service  & Coordination of asynchronous jobs\\
\hline
\end{tabular}
\label{tab:GoAnimateAmazon}
\end{table}


\vspace{5mm}

\textit{\textbf{Scenario EJT}}: \textit{easyJet} \cite{easyJet} is one of the leading European low-fare airlines. As part of its customer services, it provides an airline application for handheld devices called \textit{Halo}~\cite{easyJetAzure} featuring various electronic airline services such as check-in and seat reservation. Halo has been developed using the Windows Azure Service Bus platform \cite{WindowsAzureServiceBus} from Microsoft on which it is currently hosted \cite{easyJetAzure}. The latter resides on the Windows Azure cloud infrastructure and provides development and operational environment for scalable applications with occasionally connected clients. It features automatic management of the communication between the application users and the request processing servers handling client disconnections. The underlying Windows Azure cloud infrastructure where Halo is deployed provisions infrastructure resources to the PaaS provider Windows Azure Service Bus. This includes provisioning of, for example, networking and storage resources, that can be provided to Halo on demand in an elastic manner. For further information on the infrastructure resources offered by Windows Azure, we refer the reader to \cite{windowsazure:infrastructure}.


\vspace{5mm}

\textit{\textbf{Scenario EZS}}: \textit{EZasset} \cite{EZasset} is a business and agency asset management system delivered to end-users according to the SaaS service delivery model. Using knowledge-based technology, it efficiently manages documents and assets of a wide range of businesses. It has been built and designed to operate on the application development and hosting platform Google App Engine \cite{GoogleAppEngine} \cite{prr:ezassetlaunches}. Consequently, EZasset supports the integration of many proprietary Google APIs, such as Google Calendar and the Google Docs API \cite{ezasset:integration}. 

%

\vspace{5mm}

\textit{\textbf{Scenario FRC}}: \textit{Force.com}~\cite{Force} is a provider of platform resources for development, management, and marketing of cloud applications. Some of the central platform provisioning offerings of Force.com are \emph{Appforce} and \emph{ISVForce}. Appforce supports easy integration of cloud applications with social networks and mobile devices \cite{force:appforce}. ISVForce supports the promotion of cloud applications on the market and offers functionality to manage the process of running businesses on cloud infrastructures, including software licensing, upgrade provisioning, and similar \cite{force:isvforce}. Note that Force.com does not natively support the provisioning of infrastructure resources to end-users (i.e., application developers leasing platform resources). Since recently, Force.com users may lease infrastructure resources, but they are provisioned by Amazon as a native cloud infrastructure provider \cite{ForceForAWS}. 



\vspace{5mm}

\textit{\textbf{Scenario SFR}}: \textit{Salesforce.com} \cite{SalesForce} is a provider of CRM applications delivered to end-users according to the SaaS service delivery model. Some of its products include \emph{SalesCloud} (a set of applications for marketing and sales operations \cite{salesforce:salescloud}) and \emph{ServiceCloud} (a set of applications for customer care services \cite{salesforce:servicecloud}). Both of them include \emph{Chatter}, an application for real-time collaboration tasks. These applications rely on the platform resources provisioned by the native platform provider Force.com.
%
Note that the SaaS provider Salesforce.com and the PaaS provider Force.com are within the jurisdiction of the same company -  Salesforce.com.



\vspace{5mm}

\textit{\textbf{Scenario DNB}}: At the time of writing, \textit{DenizBank} \cite{DenizBank} is a fast growing Turkish bank in need of expansion and increased efficiency of its IT infrastructure. DenizBank has built its own private cloud infrastructure by virtualizing existing servers using the Hyper-V hypervisor technology \cite{HyperV}. It uses Microsoft System Center 2012~\cite{SystemCenter}, which enables easy and flexible management of infrastructure resources \cite{DenizBankMicrosoft}. Some of the reported benefits that DenizBank gained by building a private cloud environment include total savings of up to 19 million in data center costs and 20\% reduction of IT staff costs \cite{DenizBankMicrosoft}.


\vspace{5mm}

\textit{\textbf{Scenario ZNG}}: \textit{Zynga} \cite{Zynga} is a provider of many popular social game applications. Since its inception and shortly thereafter, Zynga has used public hosting resources in order to deploy and run gaming services. However, with the sudden immense popularity of one of their games - Farmwille - the amount and intensity of the workloads processed by Zynga's private infrastructure have drastically increased. This has been a major reason for the migration of the existing Zynga infrastructure to infrastructure resources provisioned by Amazon, i.e., Amazon Web Services \cite{AmazonWebServices} (see Scenario AWS). However, due to the need for greater control over the optimization of the infrastructure resources, Zynga has also built their own private cloud environment for provisioning of infrastructure resources. Currently most of the workloads are processed on Zynga's private infrastructure and in case the capacity of this infrastructure is saturated, Zynga's uses additional leased infrastructure resources from Amazon Web Services. Thus, Zynga's approach towards dealing with inflating workloads is a hybrid infrastructure resource provisioning for maintaining efficiency in delivering social game services to end-users \cite{techrepublic:theevolution}. 

\vspace{5mm}

\textit{\textbf{Scenario DTO}}: \textit{Dito} \cite{Dito} is an authorized Google App reseller which adds value to end-users by providing experise knowledge on the migration to software resources originally provisioned by Google. Among many other services, Dito provisions Google's software resources to end-users with additional add-on services customized according to specific end-users' needs. For instance, Dito provisions a proprietary management tool for Google Apps called \emph{Dito GAM}, which enables the efficient management of domain and user settings. For further information on the features of Dito GAM, we refer the reader to \cite{dito:ditogam}.


\subsection{Textual and Visual Cloud Usage Patterns in Practice}\label{sec:practice:patterns}




In this section, we apply our formalism for describing cloud usage patterns to the presented real-world cloud usage scenarios. The patterns describing each of the considered cloud usage scenarios are shown in Table~\ref{tab:ScenarioPatterns}. Note that we do not consider provisioning sizes/volumes since we focus on the provider-consumer relationships between the involved stakeholders. In the following, we discuss the textual form of selected patterns from Table~\ref{tab:ScenarioPatterns} relating them to the respective cloud usage scenarios:

\begin{table}[h]
\caption{cloud usage patterns that describe real-world cloud usage scenarios}
\centering
\begin{tabular}{|p{5cm}|p{8cm}|}
\hline
Cloud Usage Scenario & Cloud Usage Pattern \\
\hline 
\hline

AWS  & \texttt{i.e}\\
\hline
FBK & \texttt{nps.e}\\
\hline
GAN  & \texttt{i.s.e}\\
\hline
EJT  & \texttt{ip.s.e}\\
\hline
EZS & \texttt{p.s.e}\\
\hline
FRC & \texttt{p.e}\\
\hline
SFR & \texttt{ps.e}\\
\hline
DNB & \texttt{ie}\\
\hline
ZNG & \texttt{(i.)(i)s.e}\\
\hline
DTO & \texttt{(s.)s.e}\\
\hline
\end{tabular}
\label{tab:ScenarioPatterns}
\end{table}

\vspace{5mm}

\textbf{Scenario AWS - \texttt{i.e}:} The native cloud provider Amazon provides infrastructure resources to end-users \cite{AmazonWebServices}. In the cloud usage pattern, this is reflected by the (provider, consumer) pair (IaaS, End-user) denoted as \texttt{i.e}.  The QoS requirements and customer contracts are defined as part of an external SLA. Therefore, the letters \texttt{i} and \texttt{e} are separated by a ``dot'' (.). The hardware resources are provisioned to the intermediary at the IaaS abstraction level using virtualization technology, which is the default case. Therefore, the section \emph{Hardware resources} is omitted in the cloud usage pattern.

\vspace{5mm}

\textbf{Scenario FBK - \texttt{nps.e}:} The native cloud provider Facebook \cite{Facebook} provides a social networking application to end-users. This is reflected in the cloud usage pattern by the (provider, consumer) pair (SaaS, End-user) denoted as \texttt{s.e}. The respective QoS requirements and customer contracts are defined as part of an external SLA. Therefore, the letters \texttt{s} and \texttt{e} are separated by a ``dot'' (.). The Facebook application is deployed on platform resources provisioned by a platform provider within the jurisdiction of the same company (i.e., Facebook). This is reflected by the (provider, consumer) pair (SaaS, End-user) denoted as \texttt{ps}. The hardware resources are provisioned to the intermediary at the PaaS abstraction level without the use of virtualization technology. Therefore, the letter \texttt{n} preceding the letter \texttt{p} is included in the cloud usage pattern. 



\vspace{5mm}

\textbf{Scenario EJT - \texttt{ip.s.e}:} The non-native cloud service provider easyJet \cite{easyJet} provides a flight management application to end users. This is reflected by the (provider, consumer) pair (\emph{SaaS, End-user}) denoted as \texttt{s.e}. easyJet does not own a cloud infrastructure, but instead relies on infrastructure resources provisioned by another cloud provider, i.e., Windows Azure. The QoS requirements and customer contracts related to the provisioning of the application to end-users are defined as part of an external SLA. Therefore, the letters \texttt{s} and \texttt{e} are separated by a ``dot'' (.). The easyJet application is implemented and deployed on platform resources leased from Windows Azure~\cite{WindowsAzure}, the latter being a separate legal entity. This is reflected by the (provider, consumer) pair (PaaS, SaaS) denoted as \texttt{p.s}.  
The provisioning of platform resources is supported by infrastructure resources also provisioned by Windows Azure. This is reflected by the (provider, consumer) pair (IaaS, PaaS) denoted as \texttt{ip}. 
The hardware resources are provisioned to the consumer at the IaaS abstraction level using virtualization technology. Therefore, as usual, the \emph{Hardware resources} section is omitted and no letter precedes the letter \texttt{i}. Note that the platform provider Windows Azure is considered as a native cloud provider since it owns a cloud infrastructure; that is, the ommited \emph{Hardware resources} section and the \emph{IaaS} section of the respective cloud usage pattern are within the legal boundaries of the provider Windows Azure. 

\vspace{5mm}

\textbf{Scenario DNB - \texttt{ie}:} DenizBank \cite{DenizBank} has deployed the infrastructure provider platform Microsoft System Center 2012 for managing and scaling the bank's private IT infrastructure \cite{DenizBankMicrosoft}.
This is reflected in the pattern by the (provider, consumer) pair (IaaS, End-user) denoted as \texttt{ie}. Given that in this case, the end-users and the IaaS provider are part of the same organization, the respective QoS requirements and customer contracts are defined as part of an internal SLA and therefore, the letters \texttt{i} and \texttt{e} are not separated by a ``dot'' (.). The hardware resources are provisioned to the infrastructure provider with the use of virtualization technology \cite{DenizBankMicrosoft}. Therefore, as usual, the \emph{Hardware resources} section is omitted and no letter precedes the letter \texttt{i}. Note that this scenario is an example of a usage scenario of a private cloud, reflected by the lack of dots in the cloud usage pattern. 

\vspace{5mm}

\textbf{Scenario ZNG - \texttt{(i.)(i)s.e}:} Zynga \cite{Zynga} provides gaming applications to end-users according to the SaaS service delivery model. This is reflected by the (provider, consumer) pair (SaaS, End-user) denoted as \texttt{s.e}. For scalability and efficiency, Zynga uses provisioned infrastructure resources that it owns, but it also leases additional infrastructure resources from Amazon Web Services. To this end, the respective cloud usage pattern contains two pattern specifications enclosed in parentheses (see Rule H.II, Section~\ref{sec:patterns:textual}): \texttt{(i.)} specifying the provisioning of infrastructure resources from Amazon Web Services, and \texttt{(i)} specifying the provisioning of infrastructure resources from Zynga's private infrastructure provider. Regarding the specification \texttt{(i.)}, note that the character ``dot'' (.) is written inside the parentheses in order to indicate that the respective infrastructure provider involved in the hybrid resource provisioning (i.e., Amazon Web Services) is not within the legal domain of Zynga. 
To the contrary, the character ``dot'' (.) is omited when writing the specification \texttt{(i)}, since Zynga's private infrastructure provider is within the same legal boundaries as the SaaS provider.

\vspace{5mm}

\textbf{Scenario DTO - \texttt{(s.)s.e}:} Dito \cite{Dito} resells software resources provisioned by Google's SaaS-level provider Google Apps. Thus, the cloud usage pattern uses parentheses (see Rule M.I, Section~\ref{sec:cupsforvaluechains}), i.e., \texttt{(s.)}, to indicate the provisioning of software resources from Google Apps to the mediator Dito
A ``dot'' (.) is included inside the parentheses in order to indicate that Google's SaaS provider is not within the legal domain of Dito (i.e., the respective QoS requirements and customer contracts are defined as part of an external SLA). The specification \texttt{(s.)} is followed by the letter \texttt{s} indicating that Dito is a SaaS-level mediator that resells software resources to end-users.

\vspace{5mm}

\begin{figure}[h]
\begin{center}
\includegraphics[scale=0.58]{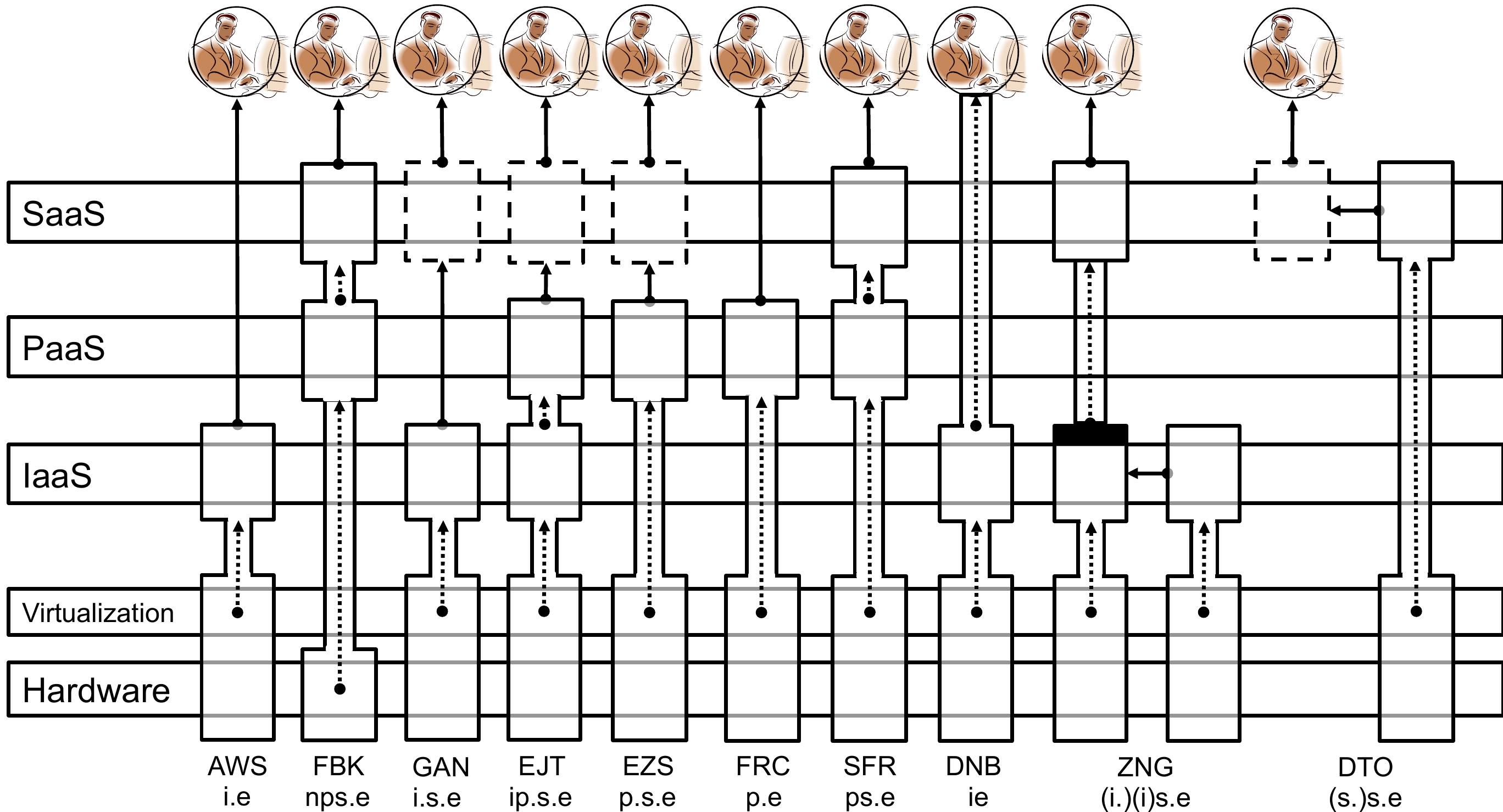} 
\end{center}
\caption{visual forms of the textual cloud usage patterns listed in Table~\ref{tab:ScenarioPatterns}
}
\label{fig:CloudPatternsScenarios}
\end{figure}

In Figure~\ref{fig:CloudPatternsScenarios}, we use the proposed visual formalism (Section~\ref{sec:patterns:visual}) to visualize the textual cloud usage patterns shown in Table~\ref{tab:ScenarioPatterns}. Note that when visualizing Scenario ZNG, we place the thick line indicating hybrid resource provisioning in the graphical element of Zynga's private IaaS provider to visualize the ability of this provider to ``spill over'' to  the IaaS provider Amazon Web Services \cite{AmazonWebServices} (i.e., to use leased infrastructure resources \cite{AmazonWebServices}).


Although in Section~\ref{sec:practice:services} and Section~\ref{sec:practice:patterns} we discussed only 10 cloud usage scenarios, there are many other real-world scenarios that conform to the presented cloud usage patterns. For instance, as in Scenario AWS, the provisioning of infrastructure resources of the popular IaaS provider Terremark \cite{terremark} to end-users conforms to the pattern \texttt{i.e}, and further, as in Scenario ZNG, the use of both leased and private infrastructure resources by the SaaS provider Music Mastermind \cite{MusicMastermind} conforms to the pattern \texttt{(i.)(i)s.e} (see \cite{edison:musicmaking}).

\subsubsection{Cloud Usage Patterns and Real-World Cloud Applications}
\label{sec:analysis-patters-applications}



In this section, we consider the set of representative real-world cloud applications selected in Section~\ref{sec:apps} to analyze the relationship between cloud usage patterns, on the one hand, and applications and application types, on the other hand. By studying which cloud usage patterns are typically used in different application domains, we reason about the advantages and disadvantages of different approaches in which cloud services can be used to support the implementation and operation of different application types. Furthermore, by studying the relationships between common cloud usage patterns and application domains we identify best practices for the adoption of cloud computing technologies in different domains. 

We first classify the applications involved in the previously presented real-world cloud usage scenarios (Section~\ref{sec:practice:services}) according to their type by means of the cloud application categorization from Section~\ref{sec:typical-apps-on-clouds}. We then map the considered applications and application types to respective cloud usage patterns. The results of this mapping are summarized in Table~\ref{tab:relationshipts-cloud-application-types-usage-patterns}. In the rest of this section, we present our observations for each of the considered applications.

\begin{table}[h]
\caption{mapping of cloud applications and application types to cloud usage patterns}
\centering
\begin{tabular}{|p{3cm}|p{6cm}|p{4.5cm}|}
\hline
Application & Application Type & Cloud Usage Pattern \\
\hline 
\hline
Facebook & Social networking & \texttt{nps.e}\\ \hline
Go!Animate & User data processing & \texttt{i.s.e}\\ \hline
EasyJet & CRM/PRM & \texttt{ip.s.e}\\ \hline
SalesForce.com &  CRM/PRM  & \texttt{ps.e}\\ \hline
Zynga &  Online gaming and meta-gaming  & \texttt{(i.)(i)s.e}\\  
\hline

\hline
\end{tabular}
\label{tab:relationshipts-cloud-application-types-usage-patterns}
\end{table}


\textbf{Facebook - \texttt{nps.e}} Facebook is a social networking application enabling the sharing and exchange of messages and multimedia among a vast number of users. Each Facebook user maintains a personalized profile that is usually updated frequently. Profile updates are propagated to related user profiles (referred to as ``friends'' in the Facebook terminology) in a continuous and timely fashion. Also, Facebook users can use platform resources provided by Facebook to develop applications, and deploy and run them on Facebook's cloud. 
Some reports indicate that Facebook's data centers in 2009 stored more than 40 billion photos, and that users upload 40 million photos each day~\cite{DataCenterKnowledgeFacebook2009}. Given the high intensity of user activities, on the one hand, and the increasingly stringent application performance requirements, on the other hand, it can be concluded that one of Facebook's main technological requirements is to support scalability and efficiency of operation. 
 This requirement is the major reason behind the cloud usage pattern (i.e., \texttt{nps.e}) used to implement Facebook's services. Gio Coglitore of Facebook Labs has stated that the lack of using virtualization technology for the provisioning and management of hardware resources, denoted by the letter \texttt{n} in the respective cloud usage pattern, significantly reduces the overhead of scaling the application during operation:

\begin{quote}
\emph{``We find within our testing that a realised [non-virtualised] environment brings efficiencies and the ability to scale much more effectively''} 

\emph{Source}: PC World Magazine, IDG News Service, March, 2011 \cite{lawson:facebook}.

\end{quote}


\textbf{Go!Animate - \texttt{i.s.e}} Go!Animate is a web application for development of amateur animation videos. A major requirement for the effective functioning of Go!Animate is the provisioning of infrastructure resources for hosting its operating systems, application components, and databases in a scalable manner~\cite{GoAnimateAmazon}. To this end, Go!Animate leases infrastructure resources from Amazon. In this direction, Go!Animate's Chief Executive Officer Alvin Hung has stated:

\begin{quote}
\emph{``It was important to put our tech stack on a flexible and cost effective architecture... The ability to bring up and shut down instances easily has given us much more flexibility...''} 

\emph{Source}: Amazon, Case Studies, February, 2011 \cite{GoAnimateAmazon}.

\end{quote}

Note that Go!Animate does not lease platform resources as indicated by the lack of the letter \texttt{p} in the respective cloud usage pattern, a major reason being the fact that Go!Animate requires maximum flexibility at the platform level (i.e., its operation relies on a set of custom platform components developed specifically for the Go!Animate application, as opposed to using readily available platform services offered by cloud providers).


\textbf{EasyJet - \texttt{ip.s.e}} The EasyJet's Halo application is a CRM application. Similar to other applications of this type, Halo features complex business logic which, on the one hand, requires many common enterprise middleware services (e.g., transaction processing, data persistence, access control, and fault tolerance mechanisms), and, on the other hand, dynamic resource provisioning and elastic capacity management. This calls for the use of platform resources that are supported by provisioning of underlying infrastructure resources, which Halo obtains from the Windows Azure cloud environment acting as an infrastructure and platform provider at the same time. 
Regarding the benefits of leveraging both infrastructure and platform resource provisioning according to the IaaS and PaaS service delivery models, EasyJet's Enterprise Architect Bert Craven has stated:

\begin{quote}
\emph{``We don't have to build a new high-availability service platform, make firewall configuration changes, or deploy lots of new servers. From the service consumer's point of view, there is no difference in how they get to that service.''} 

\emph{Source}: Microsoft, Case Studies, August, 2011 \cite{easyJetAzure}.

\end{quote}





\textbf{Salesforce.com - \texttt{ps.e}/ Force.com - \texttt{p.e}} The development and hosting of the Salesforce.com CRM application \cite{SalesForce} is supported by platform resources provisioned by the platform provider Force.com (Scenario FRC, Section~\ref{sec:practice:services}). The latter enables rapid application development by providing support for many essential platform-level features such as the integration with various APIs (e.g., Google Apps, web services APIs), and a developer sandbox. 
%
The platform services of Force.com have recently been extended to support the use of provisioned infrastructure resources from Amazon, resulting in a new product called Force.com for Amazon Web Services~\cite{ForceForAWS}. At the time of writing, we do not have any evidence that infrastructure resources from Amazon are intended to also be used for the Salesforce.com applications. If that were to happen in the future, the Salesforce.com applications would conform to the pattern \texttt{i.ps.e}. 
Force.com for Amazon Web Services has been positively received by many Force.com customers, including John Johnson, Vice President of Licensing at ASCAP, who has stated:

\begin{quote}
\emph{``We're excited to use Force.com and Amazon Web Services together to run our business in the cloud. We've been able to leverage these cloud services to create new applications, including document management, that bring new efficiencies to our business.''} 

\emph{Source}: Salesforce.com, Press Releases, November, 2008 \cite{SalesForceAmazonWS}.

\end{quote}

\textbf{Zynga - \texttt{(i.)(i)s.e}} As we previously mentioned (Section~\ref{sec:practice:services}), Zynga \cite{Zynga} has initially relied solely on infrastructure resources provided by Amazon \cite{AmazonWebServices} in order to maintain efficiency in provisioning social gaming services to end-users. However, due to the need for more fine-granular management and customization of the infrastructure resources, Zynga has decided to use a hybrid resource provisioning approach; that is, Zynga has built its own infrastructure cloud provider and in addition, it leases additional infrastructure resources from Amazon when the capacity of Zynga's private infrastructure provider is saturated. In this direction, Allan Leinward, a Chief Technology Officer of Infrastructure for Zynga, has stated:

\begin{quote}
\emph{``...we came to the realization that we were renting what we could own. The public cloud isn’t your own infrastructure; it isn’t something you can own and operate in your own way, and it isn’t capital equipment, so it was an operating expense.''} 

\emph{Source}: TechRepublic, Blog Entry, March, 2012 \cite{techrepublic:theevolution}.

\end{quote}

%


\newpage 

\section{Related Work}\label{sec:related}

Our work closely relates to several research topics including: (i) analysis of cloud usage trends, characteristics of cloud environments, and challenges in using these environments, (ii) classification of cloud applications, and (iii) construction of formalisms for expressing patterns. In this section, we review relevant related work focusing on these research topics.

\subsection{Cloud Computing Characteristics, Challenges, and Use Case Scenarios}\label{sec:related:challenges}

Armbrust et al. \cite{Armbrust+09} provide an in-depth analysis of cloud environments; that is, they provide an overview of many concerns related to cloud environments, such as conrete and accurate definition of these environments, identification of their unique characteristics, identification of representative application types, and similar. They identify the following features and characteristics that distinguish cloud environments from their traditional counterparts: (i) on-demand provisioning of ``infinite'' computing resources, (ii) scalability, (iii) pay-as-you-go business model. Armbrust et al. \cite{Armbrust+09} also identify and analyze application types commonly seen in cloud environments such as mobile interactive applications, parallel batch processing applications, compute-intensive desktop applications, and so on. Further, Armbrust et al. \cite{Armbrust+09} study the cloud computing market at the time of writing, analyzing the popular cloud computing services Amazon Web Services \cite{AmazonWebServices}, Microsoft Azure \cite{WindowsAzure}, Google AppEngine \cite{GoogleAppEngine}, and others. This includes observations of the effects of elasticity, as a unique cloud computing characteristic, on the revenue and the expenses of cloud providers, on the economical gains to customers of these providers, and so on. Finally, Armbrust et al. \cite{Armbrust+09} analyze some of the major challenges in cloud computing such as security concerns, not standardized APIs for development of cloud applications, and similar. They also provide an overview of promising future trends such as the development of software for deployment in virtualized environments, development of novel billing practices, standardization of cloud technologies, and many other. 

Another related work with a focus on the analysis of cloud technologies and usage scenarios is \cite{OSGCloud12} by the SPEC OSG Cloud Working Group of SPEC. OSG \cite{OSGCloud12} provides an extensive analysis of the challenges and requirements in the area of performance benchmarking of cloud environments. Based on earlier work by Mell et al. \cite{mell+2010}, OSG \cite{OSGCloud12} characterizes a given environment as a cloud environment if it has the following characteristics: on-demand self-service, broad network access, resource pooling, rapid elasticity, and measured service. For a detailed description of these characteristics, we refer the reader to \cite{OSGCloud12} and \cite{mell+2010}. Because of the relevance to cloud performance benchmarking, OSG \cite{OSGCloud12} analyzes typical cloud usage scenarios such as social networking, data analytics, and voice-over-IP.  They also relate the identified usage scenarios to specific organizations that implement such scenarios, for example, Facebook \cite{Facebook} as a provider of social networking services. Also, OSG \cite{OSGCloud12} defines a cloud SUT as a system consisting of a set of components (e.g., services, hardware, and software components) that are exercised by a workload representative for workloads in cloud environments. OSG \cite{OSGCloud12} also defines a FDR (Full Disclosure Report) as a report containing a detailed description of the components of a given cloud SUT. Such a report is important since it enables a benchmarker to verify that a cloud SUT (system under test) conforms to a given set of benchmarking run rules, and further, it enables the reproduction of benchmarking experiments. Among the many other contributions of the work of OSG \cite{OSGCloud12} are an analysis of the unique characteristics of a cloud benchmark, a specification of relevant metrics (e.g., elasticity, throughput, power, price), and a survey of existing cloud evaluation tools and frameworks (e.g., the cloud benchmarking frameworks proposed by Cooper et al. \cite{Cooper+10} and Sobel et al. \cite{Sobel+2008}). 

Amrhein et al. \cite{amrhein:cloud} leverage the previously mentioned definitions of cloud computing and cloud computing services by Mell et al. \cite{mell+2010} (dated 8-19-09; Section~\ref{sec:terms-and-definitions}) to identify and characterize common cloud use case scenarios. This includes end-user to cloud (i.e., a scenario in which an end-user accesses data or applications in the cloud), enterprise to cloud to end-user (i.e., a scenario in which an enterprise delivers cloud services to an end-user, the end-user being someone within the enterprise, or an external user), enterprise to cloud to enterprise (i.e., a scenario in which resources are hosted on a cloud such that multiple enterprises can interoperate), and so on. Amrhein et al. \cite{amrhein:cloud} also identify relevant requirements related to each of the previously mentioned use case scenarios, for example, SLAs, security, identity, and so on. In addition, Amrhein et al. \cite{amrhein:cloud} describe customers' experiences with each of the identified cloud usage scenarios, considering customers such as central and local government, space agencies executing astronomic data processing, and so on, identifying solved problems of the considered customers by the migration to cloud environments.

\subsection{Cloud Applications}\label{sec:related:apps}

In Section~\ref{sec:apps}, we categorized many applications that are typically deployed in cloud environments, and selected a smaller set of representative cloud applications for use throughout this work when discussing common cloud usage scenarios. In this section, we briefly review related work on application taxonomies and on the identification of representative cloud applications. 

Categorizing computer software applications has a long history that has co-existed and has developed in parallel to the evolution of software itself. Glass and Vessey \cite{Glass1992} provide an overview of the history of software categorization by reviewing taxonomies of software domains published between 1955 and 1992. A more recent taxonomy than those surveyed by Glass and Vessey \cite{Glass1992} is the taxonomy proposed by Forward and Lethbridge \cite{Forward2008} (2008) that consists of approximately 200 categories. They divide the software space into four main categories: data-dominant software, systems software, control-dominant software, and computation-dominant software, where each category has multiple sub-categories. As an illustration, in Figure~\ref{fig:general-soft-taxonomy}, we depict the top two category levels of the Forward and Lethbridge taxonomy \cite{Forward2008}. 

\begin{figure}[ht]
\begin{multicols}{2}
\textbf{A} Data-dominant software \\
--- \textbf{A.con} Consumer-oriented software \\
--- \textbf{A.bus} Business-oriented software \\
--- \textbf{A.des} Design and engineering software \\
--- \textbf{A.inf} Information display and transaction entry \\
\textbf{B} Systems software \\
--- \textbf{B.os} Operating systems \\
--- \textbf{B.net} Networking / Communications \\ 
--- \textbf{B.dev} Device / Peripheral drivers \\
--- \textbf{B.ut} Support utilities \\
--- \textbf{B.mid} Middleware and system components \\ 
--- \textbf{B.bp} Software Backplanes (e.g. Eclipse) \\ 
--- \textbf{B.svr} Servers \\
--- \textbf{B.mal} Malware \\
\textbf{C} Control-dominant software \\
--- \textbf{C.hw} Hardware control \\
--- \textbf{C.em} Embedded software \\
--- \textbf{C.rt} Real time control software \\
--- \textbf{C.pc} Process control software (i.e. air traffic control, industrial process) \\
\textbf{D} Computation-dominant software \\
--- \textbf{D.or} Operations research \\
--- \textbf{D.im} Information management and manipulation \\ 
--- \textbf{D.art} Artistic creativity \\
--- \textbf{D.sci} Scientific software \\
--- \textbf{D.ai} Artificial intelligence \\
\end{multicols}
\caption{the top two category levels of the Forward and Lethbridge taxonomy \cite{Forward2008}}
\label{fig:general-soft-taxonomy}
\end{figure}

Regarding cloud applications in particular, a survey conducted by the IBM Institute for Business Value \cite{IBM-cloud-survey-wp} in 2009 presented a selection of application types commonly seen in cloud environments. This survey focuses on applications that have been migrated to, or have been natively developed in, cloud environments, and identifies six classes of such applications, which we depict in Figure~\ref{fig:ibm-workload-taxonomy}. We extended thus survey by including several additional relevant application categories (e.g., gaming and e-Science applications) in the context of our analysis of common cloud usage scenarios.

\begin{figure}[ht]
\begin{multicols}{2}
\textbf{Analytics}  \\
--- Data mining, text mining or other analytics \\
--- Data warehouses or data marts \\
--- Transactional databases \\
\textbf{Business services}  \\
--- CRM or sales force automation \\
--- E-mail \\
--- ERP applications \\
--- Industry-specific applications \\
\textbf{Collaboration}  \\
--- Audio/Video/Web conferencing \\
--- Unified communications \\
--- VoIP infrastructure \\

\pagebreak\textbf{Desktop and devices}  \\
--- Desktop \\
\textbf{Development and test}  \\
--- Development environment \\
--- Test environment \\
\textbf{Infrastructure}  \\
--- Application servers \\
--- Application streaming \\
--- Business continuity/disaster recovery  \\
--- Data backup and archiving \\
--- Data center network capacity \\
--- Security, servers, storage \\
--- Training infrastructure \\
--- WAN capacity \\
\end{multicols}
\caption{categories of representative cloud applications as identified by the IBM Institute for Business Value \cite{IBM-cloud-survey-wp}}
\label{fig:ibm-workload-taxonomy}
\end{figure}

\subsection{Cloud Usage Taxonomies}\label{sec:related:usage}

The research and industrial communities have developed multiple taxonomies that categorize the cloud computing space with respect to different cloud usage scenarios. For instance, Intel \cite{IntelWP10a} has built a taxonomy for the purpose of establishing a common cloud computing terminology for use across the company. The proposed taxonomy segments the cloud space according to the different types of resources provisioned to customers, i.e., according to the service delivery models \emph{Software-as-a-Service}, \emph{Platform-as-a-Service}, \emph{Infrastructure-as-a-Service}, \emph{Service-as-a-Service}, \emph{Client Software}, and \emph{Cloud Client}. Intel \cite{IntelWP10a} focuses on the enterprise aspects of the common service delivery models SaaS, PaaS and IaaS, also identifying additional categories such as \emph{Service-as-a-Service}, \emph{Client Software}, and \emph{Cloud Client}. The \emph{Service-as-a-Service} category represents auxiliary services to the services delivered as part of the \emph{SaaS}, \emph{PaaS}, and \emph{IaaS} delivery models (e.g., monitoring, accounting and billing). The \emph{Cloud Software} category represents software products developed for deployment in cloud environments (e.g., data software, computing software). This category classifies software products according to the cloud properties that are relevant for Independent Software Vendors (ISV), the latter being considered as an important part of the cloud market. Finally, the \emph{Cloud Client} category represents client-centric services that directly impact the customers' experience, for example, widgets, context awareness services, and client application services. 

Similar to Intel \cite{IntelWP10a}, IBM \cite{ibm:defining} has developed a taxonomy for cloud computing, consisting of the categories \emph{cloud delivery models} (i.e., a \emph{private}, \emph{public}, and \emph{hybrid} cloud, each consisting of several sub-types such as \emph{exploratory} and \emph{departmental} cloud - sub-types of a private cloud, \emph{exclusive} and \emph{open} cloud - sub-types of a public cloud, and so on), \emph{cloud service types} (i.e., \emph{infrastructure}, \emph{platform}, and \emph{application} cloud services, and \emph{roles in cloud consumption and delivery} (i.e., \emph{consumers}, \emph{providers}, and \emph{integrators}). Leimeister et al. \cite{leimeister+2010} also analyze the use of cloud systems from a business perspective;  they focus on structuring and conceptualizing value networks in the context of cloud computing. Under value network, they understand a network of multiple actors, i.e., suppliers of services and distributers linked together in order to create or add value for end-users. Leimeister et al. \cite{leimeister+2010} state that the main characteristic of a value network is the large amount of exchanges, interactions, and value flows between the different actors. Leimeister et al. \cite{leimeister+2010} classify these actors into \emph{customers} (i.e., buyers of services through various distribution channels), \emph{service providers} (i.e., developers and operators of services adding value to customers), \emph{consultants} (i.e., providers of support to customers in the selection and implementation of relevant services), and so on. For a detailed description of the different actors, we refer the reader to \cite{leimeister+2010}. Leimeister et al. \cite{leimeister+2010} also demonstrate the use of the e$^3$-value method \cite{Gordijn02} for the visualization of value networks. This method enables visual presentation of economically valuable objects in a value network consisting of multiple actors. 

In contrast to the taxonomies used in industry, taxonomies developed by the research community are less focused on the enterprise aspects of cloud systems. For instance, Oliviera et al. \cite{Oliveira+10} segment the cloud space according to the needs of researchers for using cloud environments for scientific experimentation. They define eight top-level categories: \emph{architecture}, \emph{business model}, \emph{technology infrastructure}, \emph{privacy}, \emph{standards}, \emph{pricing}, \emph{orientation}, and \emph{access}. Oliviera et al. \cite{Oliveira+10} also propose sub-categories, for example, the \emph{technology infrastructure} category classifies cloud systems into systems \emph{with HPC Support} and systems \emph{without HPC Support}, where the former is considered as an important feature for conducting computationally intensive scientific experiments. The three classical cloud service delivery models (i.e., \emph{SaaS}, \emph{PaaS} and \emph{IaaS}) belong to the \emph{business model} category as part of which an additional category \emph{Storage-as-a-Service} (with \emph{Database-as-a-Service} as a sub-category) is defined. The latter is used to explicitly distinguish data management and storage services, which are crucial for managing and storing results of scientific experiments. Oliviera et al. \cite{Oliveira+10} use the proposed taxonomy to classify existing popular cloud environments such as Amazon EC2 \cite{AmazonEC2} and Windows Azure \cite{WindowsAzure}.

Other taxonomies focusing on specific uses of cloud environments are proposed by Abbadi et al. \cite{Abbadi11}, Rimal et al. \cite{Rimal+10}, and Foster et al. \cite{foster+2009}. Abbadi et al. \cite{Abbadi11} propose a taxonomy classifying the components of cloud environments that are relevant for the development and operation of autonomic cloud management services maintaining reliable and efficient operation of cloud applications. Abbadi et al. \cite{Abbadi11} classify the different components of cloud environments according to: (i) their nature, as \emph{physical}, \emph{virtual}, or \emph{application} components, and (ii) their function, as \emph{server}, \emph{network}, or \emph{storage} components. Abbadi et al. \cite{Abbadi11} ``slice'' the cloud space into a \emph{horizontal} and \emph{vertical} dimension, the former structuring the cloud space in \emph{physical}, \emph{virtual} and \emph{application} layers, the latter further dividing each layer into \emph{server}, \emph{network}, and \emph{storage} sub-layers. As a use case scenario, Abbadi et al. \cite{Abbadi11} focus on analysing the requirements for deployment and operation of a multi-tier weather forecast application in a cloud environment. Abbadi et al. \cite{Abbadi11} map required cloud infrastructure components to respective taxonomy categories in order to identify relevant autonomic cloud management services for management of these components so that the considered weather forecast application operates efficiently.

Rimal et al. \cite{Rimal+10} analyze the cloud space in terms of features of cloud environments relevant for massive data processing applications. Rimal et al. \cite{Rimal+10} propose the categories \emph{cloud architecture}, \emph{virtualization management}, \emph{service}, \emph{fault tolerance}, \emph{security}, and \emph{other}. As part of the \emph{cloud architecture} category, Rimal et al. \cite{Rimal+10} define the sub-categories \emph{private}, \emph{public}, and \emph{hybrid} clouds representing different data access restrictions in cloud environments. The category \emph{fault tolerance} classifies fault tolerance mechanisms that may be implemented in cloud environments. As part of the category \emph{other} sub-categories such as \emph{load balancing}, \emph{interoperability}, \emph{scalable data storage}, and so on, are defined. Rimal et al. \cite{Rimal+10} further segment the sub-category \emph{scalable data storage} into \emph{vertical} and \emph{horizontal} data storage scalability. Rimal et al. \cite{Rimal+10} use their taxonomy to analyze popular cloud service providers such as Amazon Web Services \cite{AmazonWebServices}, GoGrid \cite{GoGrid}, and Flexiscale \cite{FlexiScale}.

Foster et al. \cite{foster+2009} focus on comparing cloud computing and grid computing by discussing similarities and differences in the features and use of these computing paradigms. To this end, Foster et al. \cite{foster+2009} structure the unique characteristics of cloud computing into six categories, i.e., \emph{business model}, \emph{architecture}, \emph{resource management}, \emph{programming model}, \emph{application model}, and \emph{security model}. As an example, when considering the category business model, Foster et al. \cite{foster+2009} state that cloud computing features billing by considering resource utilization, whereas the payment for grid computing resources is project-oriented (i.e., customers pay for the use of grid resources for a fixed amount of time estimated to be sufficient for completing the relevant project tasks). We refer the reader to \cite{foster+2009} for details on these categories. Foster et al. \cite{foster+2009} also provide an outlook on future activities in the domain of cloud computing that would enhance the efficiency and usability of cloud environments, such as the development of resource provisioning methods precisely following the customers' demands, the development of protocols for close monitoring and management of resource reservations, and so on.

Some taxonomies, such as the ones proposed by Prodan et al. \cite{Prodan+09}, Lenk et al. \cite{lenk:whats}, Hofer et al. \cite{HoferK2011}, and Liu et al. \cite{liu:nist}, do not focus on a specific use of cloud environments, but instead categorize cloud environments from a general perspective. Prodan et al. \cite{Prodan+09} survey 14 cloud environments and classify them into the categories \emph{service type}, \emph{resource deployment}, \emph{hardware}, \emph{runtime tuning}, \emph{security}, \emph{business model}, \emph{middleware}, and \emph{performance}. As part of these categories, they define many further interrelated sub-categories. For instance, the \emph{SaaS}, \emph{PaaS}, and \emph{IaaS} service delivery models are defined as sub-categories of the \emph{service type} category. Also, a sub-category of \emph{service type} is \emph{specialized services}, which is used for classifying services into \emph{web hosting} and \emph{file hosting} services. The work of Lenk et al. \cite{lenk:whats} focuses on a categorization classifying known cloud services into the main service categories \emph{IaaS}, \emph{PaaS}, and \emph{SaaS}. They define sub-categories such as \emph{Physical Resource Set Services} and \emph{Virtual Resource Set Services} (parts of the \emph{IaaS} category), \emph{Programming Environments} and \emph{Execution Environments} (parts of the \emph{PaaS} category), and \emph{Basic Application Services} and \emph{Composite Application Services} (parts of \emph{SaaS} category).

Hofer et al. \cite{HoferK2011} construct a taxonomy that segments the cloud space according to several relevant characteristics of cloud computing services such as \emph{license type} (i.e., proprietary, open-source), \emph{intended user group} (i.e., corporate, private), \emph{security and privacy} (e.g., encryption, authentication), and others. Hofer et al. \cite{HoferK2011} also define the category \emph{openness of clouds} classifying cloud environments according to the amount of available information on the provisioned cloud services to costumers. This includes information on the used hardware and software, the deployed security solutions, and similar. As part of the category \emph{openness of clouds}, Hofer et al. \cite{HoferK2011} define the sub-categories \emph{unknown/limited}, \emph{basic}, \emph{moderate}, and \emph{complete}. Further, they build upon the existing common categorization of cloud service delivery models as SaaS, PaaS, and IaaS, and define additional sub-categories to reflect specific characteristics of each service delivery model. We refer the reader to \cite{HoferK2011} for further details on these sub-categories. Finally, Hofer et al. \cite{HoferK2011} classify the cloud computing services of several popular cloud providers (e.g., Amazon EC2 \cite{AmazonEC2}, Microsoft Azure \cite{WindowsAzure}, and Google Apps \cite{GoogleAppEngine}). 

Liu et al. \cite{liu:nist} propose a four-level taxonomy that describes major actors as well as their roles and responsibilities and consists of the levels: \emph{role} (i.e., set of obligations and behaviors associated with actors), \emph{activity} (i.e., behaviors/tasks associated with a given role), \emph{component} (i.e., process, actions, or tasks needed for meeting the objective of a given activity), and \emph{sub-component} (i.e., a modular part of a component). Similar to Leimeister et al. \cite{leimeister+2010}, Liu et al.  \cite{liu:nist} identify the actors \emph{Cloud Consumer}, \emph{Cloud Provider}, \emph{Cloud Broker}, \emph{Cloud Auditor}, and \emph{Cloud Carrier}. We refer the reader to \cite{liu:nist} for further details on this taxonomy.

\subsection{General Pattern Languages and Formalisms}\label{sec:related:general}

Over the past half-century, a variety of pattern languages and formalisms have been developed to ease discussions and design in many areas of computer science and engineering, and as such, they have proven influential for many application domains. In this section, we present a non-exhaustive survey of such pattern languages and formalisms. In contrast to the surveyed body of work, our work focuses on applications that are built on top of cloud computing services.

In computer science and information theory, Shannon and Weaver \cite{book/ShannonW49} formalized and analyzed general communication systems. Parnas~\cite{DBLP:journals/tse/Parnas76} was one of the first to formalize the idea of design patterns in software engineering in particular; two decades later, the work of Gamma et al. (also known as ``the gang of four'') on code design patterns \cite{book/GammaHJV94} identified many elemental patterns that are useful in the development of practical software packages. Further, Blaauw and Brooks~\cite{book/BlaauwB97} introduced a formalism for the description of computer architectures and used it to describe a variety of practical computer designs. Finally, De Marco et al.~\cite{book/DeMarco08} analyzed project behavior in IT projects by developing and using project behavioral patterns.

In parallel with developments in the computer science domain, work in linguistics, architecture, and other scientific domains has also led to the design and development of patterns and formalisms. These works have proven valuable for decades facilitating the discussion about, and design around, the patterns and formalisms that they introduce. For instance, Propp~\cite{book/Propp68} introduced a structural formalism that has been used for matching a large number of Russian folk tales to global mythos and folk tales. This formalism was later adapted by Campbell~\cite{book/Campbell72} and by Vogler~\cite{book/Vogler07}, works still used in practice, for example, for the development of Hollywood scripts~\cite{book/McKee99}. In the early 1960s, and in a more refined form at the end of the 1970s, Alexander~\cite{book/Alexander64} introduced a formalism for describing architectural elements and usage patterns in design processes in general; despite the initial controversy, it is widely considered as a milestone in the development of architectures~\cite[Chapter 1]{book/Brooks10}. These parallel developments have at times found their way into computer science, for example, the work of Smith and Williams \cite{smith:performance}, based on the work of Alexander~\cite{book/Alexander64}, provides a collection of software performance patterns and anti-patterns.

\newpage 

\section{Conclusion}
\label{sec:conclusions}

Contrasting the growth of cloud computing as an important branch of automated ICT services, discussing the elements involved in a cloud-based ICT scenario is still relying on imprecise terminology and ambiguously used concepts. To address this situation, previous work has focused on defining and formalizing general types of service delivery models (i.e., IaaS, PaaS, and SaaS). To the contrary, in this work we proposed a formalism for describing common real-world cloud usage scenarios, referred to as cloud usage patterns.

Our formalism takes a structuralist approach by defining individual elements of composition that correspond to the typical cloud service models and to various other aspects of cloud usage, such as virtualization, service leasing, and service composition. We specifically designed our formalism to support several cloud usage patterns that have emerged recently, such as hybrid services and value chains. We also proposed simple yet expressive textual and visual languages for our formalism. 

We designed our formalism to support both abstract usage, for example, in the design of cloud services through the innovative composition of existing elements, and practical usage, for example, in the discussion between benchmarking professionals through the precise description of benchmarking scenarios. By decomposing a cloud usage scenario into individual elements corresponding to the common cloud service delivery models and by describing each of them using our formalism, a computer expert will be able to communicate to a wider audience or to a prospective customer. By expressing multiple cloud usage scenarios with our formalism, a designer will be able to extract common usage features and express them with cloud used patterns, which can then be shared with a community of experts and users.

We showed comprehensive evidence that our formalism can be used in practice for describing a variety of cloud usage scenarios derived from real-world situations. Among the cloud scenarios for which we demonstrated our formalism are: global infrastructure providers leasing resources to end-users, companies offering online customer and ticketing services with software hosted entirely on external cloud platforms, companies offering online asset management and banking services while also managing a private cloud infrastructure, companies providing online social gaming services while also leasing resources from multiple cloud providers, and so on. As part of our future work, we plan to extend our formalism to support further emerging cloud service delivery models, such as Data-as-a-Service and Science-as-a-Service. 
\newpage
\section*{Glossary}
\label{sec:glossary}
\addcontentsline{toc}{section}{Glossary}
\markboth{{Glossary}}{Glossary}

\hspace{0.5cm} \textbf{Resources}

\vspace{0.5cm}
\emph{Infrastructure resources}: Processing, network, storage, and other fundamental computing resources that enable the deployment and running of arbitrary software, including operating systems and applications.

\vspace{0.5cm}
\emph{Platform resources}: Tools, libraries, software development, deployment, testing and hosting environments that enable software configuration management, database integration, state management, application versioning, and so on. 

\vspace{0.5cm}
\emph{Software resources}: Software applications.

\vspace{0.5cm}
\vspace{0.5cm}
\textbf{Systems}

\vspace{0.5cm}
\emph{Cloud system} (synonyms: cloud infrastructure, cloud platform, cloud environment, cloud): Collection of hardware and software that enables the five essential characteristics of cloud computing: on-demand self-service, broad network access, resource pooling, rapid elasticity, and measured service. 

\texttt{On-demand self-service}: A consumer can unilaterally provision computing capabilities, such as server time and network storage, as needed automatically without requiring human interaction with each service provider. (Quoted from Mell et al. \cite{mell+2010})

	\texttt{Broad network access}: Capabilities are available over the network and accessed through standard mechanisms that promote use by heterogeneous thin or thick client platforms (e.g., mobile phones, tablets, laptops, and workstations). (Quoted from Mell et al. \cite{mell+2010})

	\texttt{Resource pooling}: The provider’s computing resources are pooled to serve multiple consumers using a multi-tenant model, with different physical and virtual resources dynamically assigned and reassigned according to consumer demand. There is a sense of location independence in that the customer generally has no control or knowledge over the exact location of the provided resources but may be able to specify location at a higher level of abstraction (e.g., country, state, or datacenter). Examples of resources include storage, processing, memory, and network bandwidth. (Quoted from Mell et al. \cite{mell+2010})

	\texttt{Rapid elasticity}: Capabilities can be elastically provisioned and released, in some cases automatically, to scale rapidly outward and inward commensurate with demand. To the consumer, the capabilities available for provisioning often appear to be unlimited and can be appropriated in any quantity at any time. (Quoted from Mell et al. \cite{mell+2010})

	\texttt{Measured service}: Cloud systems automatically control and optimize resource use by leveraging a metering capability at some level of abstraction appropriate to the type of service (e.g., storage, processing, bandwidth, and active user accounts). Resource usage can be monitored, controlled, and reported, providing transparency for both the provider and consumer of the utilized service. (Quoted from Mell et al. \cite{mell+2010})

\newpage
\vspace{0.5cm}
\vspace{0.5cm}
\textbf{Providers}

\vspace{0.5cm}
\emph{Infrastructure provider} (synonyms: infrastructure service provider, infrastructure-level provider, IaaS provider): An organization, i.e., a legal entity, that provides infrastructure resources to consumers according to the IaaS service delivery model.

\vspace{0.5cm}
\emph{Platform provider} (synonyms: platform service provider, platform-level provider, PaaS provider): An organization, i.e., a legal entity, that provides platform resources to consumers according to the PaaS service delivery model.

\vspace{0.5cm}
\emph{Software provider} (synonyms: software service provider, software-level provider, SaaS provider): An organization, i.e., a legal entity, that provides software resources to consumers according to the SaaS service delivery model.

\vspace{0.5cm}
\emph{Cloud provider} (synonym: cloud service provider): An organization, i.e., a legal entity, that acts as infrastructure, platform, and/or software provider.

\cleardoublepage
\pagenumbering{gobble}
\pagestyle{empty}
\renewcommand\bibname{References}

\bibliographystyle{plain}
\bibliography{SPEC-RG-2013-001_CloudUsagePatterns}

\end{document}